\documentclass[10pt]{amsart}
\usepackage{amsmath, amssymb, latexsym, enumerate}

\newcommand{\Id}{{\mathrm{Id}}}
\newcommand{\pvee}{{\scriptstyle\vee}}

\newcommand{\ee}{{\mathrm{e}}}
\newcommand{\dd}{{\mathrm{d}}}
\newcommand{\barT}{\overline{T}}
\newcommand{\bara}{\bar{a}}
\newcommand{\bart}{\bar{t}}

\newcommand{\counit}{\varepsilon}
\newcommand{\counitT}{\epsilon}

\newcommand{\counitB}{\varepsilon_{\scriptscriptstyle \mathcal{B}}}

\newcommand{\deltaB}{\delta_{\scriptscriptstyle \mathcal{B}}}
\newcommand{\grad}{{\mathrm{deg}}}

\newcommand{\Bcal}{\ensuremath{\mbox{$\mathcal B$}}}
\newcommand{\calD}{\ensuremath{\mbox{$\mathcal D$}}}
\newcommand{\Ccal}{\ensuremath{\mbox{$\mathcal C$}}}

\newcommand{\Fcal}{\ensuremath{\mbox{$\mathcal F$}}}
\newcommand{\Mcal}{\ensuremath{\mbox{$\mathcal M$}}}

\newcommand{\action}{\triangleright}

\newcommand{\ix}[1]{{}_{\scriptscriptstyle(#1)}}
\newcommand{\potimes}{{\circ}}

\newcommand{\tb}{\ensuremath{\mbox{$T(\Bcal)$}}}
\newcommand{\tbk}[1]{\ensuremath{\mbox{$T^{#1}(\Bcal)$}}}
\newcommand{\tbn}{\ensuremath{\mbox{$T^n(\Bcal)$}}}
\newcommand{\ttb}{\ensuremath{\mbox{$T(T(\Bcal)^+)$}}}
\newcommand{\ttc}{\ensuremath{\mbox{$T(T(\Ccal)^+)$}}}
\newcommand{\sbcal}{\ensuremath{\mbox{$S(\Bcal)$}}}
\newcommand{\ssb}{\ensuremath{\mbox{$S(S(\Bcal)^+)$}}}
\newcommand{\stb}{\ensuremath{\mbox{$S(T(\Bcal)^+)$}}}
\newcommand{\tsb}{\ensuremath{\mbox{$T(S(\Bcal)^+)$}}}

\newcommand{\sub}[1]{_{\mbox{${\scriptscriptstyle #1}$}}}
\newcommand{\bC}{\mathbb{C}}
\newcommand{\ii}[1]{{}_{\scriptscriptstyle[#1]}}
\newcommand{\act}{\triangleright}
\newcommand{\sgn}{{\mathrm{sgn}}}

\newcommand{\mub}{\ensuremath{\mu_{\scriptscriptstyle\mathcal B}}}

 \newtheorem{thm}{Theorem}
\newtheorem{pro}{Proposition}

\newtheorem{lem}{Lemma}
\newtheorem{exa}{Example}

\begin{document} 
\title[Renormalization as a functor on bialgebras]
{Renormalization as a functor on bialgebras}

\author{Christian Brouder
and William Schmitt}
\date{\today}
\address{Institut de Min\'eralogie et de Physique des Milieux
  Condens\'es, CNRS UMR7590,
 Universit\'es Paris 6 et 7, IPGP, 140 rue de Lourmel,
  75015 Paris, France. E-mail: christian.brouder@impmc.jussieu.fr}

\address{Department of Mathematics,
         The George Washington University, 1922 F St. NW,
         Washington, DC 20052, USA. E-mail: wschmitt@gwu.edu}


\begin{abstract}
The Hopf algebra of renormalization in quantum field
theory is described at a general level. The
products of fields at a point are assumed to form a
bialgebra $\Bcal$ and renormalization 
endows \ttb, the double tensor algebra of $\Bcal$,
with the structure of a noncommutative bialgebra.
When the bialgebra $\Bcal$ is commutative,
renormalization turns \ssb, the double symmetric
algebra of $\Bcal$, into a commutative 
bialgebra.
The usual Hopf algebra of renormalization is
recovered when the elements of $S^1(\Bcal)$
are not renormalized, i.e., when Feynman diagrams
containing one single vertex are not renormalized.
When $\Bcal$ is the Hopf algebra of a commutative group,
a homomorphism is established between the bialgebra
\ssb\ and the Fa\`a di Bruno bialgebra of composition
of series. The relation with the Connes-Moscovici
Hopf algebra is given.
Finally, the bialgebra \ssb\ is shown to give the
same results as the standard renormalization procedure
for the scalar field.
\end{abstract}
\bibliographystyle{amsplain}

\maketitle

\section{Introduction}
The algebraic structure of quantum fields has been thoroughly studied
but, until recently, their natural coalgebraic structure has not been 
exploited.
In references \cite{BrouderGroup24,BrouderQG,BrouderOecklI}
we used the coalgebraic structure of quantum fields to show that quantum 
groups and Hopf algebras provide an interesting tool for quantum field
theory calculations. In \cite{BrouderOecklI}, 
the relation between quantum groups and free scalar
fields was presented at an elementary level. In 
\cite{BrouderGroup24,BrouderQG}, quantum groups were employed to 
calculate interacting
quantum fields and the coalgebra structure of quantum
fields was used to derive general expressions
for the time-ordered and operator products.
Moreover the cohomology theory of Hopf algebras
was found useful to handle time-ordered products.
In the present paper, the renormalization of time-ordered products
is described in detail. 

In \cite{CKI}, Connes and Kreimer defined a Hopf algebra
on Feynman diagrams that describes the renormalization of
quantum field theory\footnote[1]{The Hopf algebra of Connes-Kreimer 
belongs to a class of Hopf algebras investigated by
Schmitt in 1994 \cite{Schmitt94}.}.
A little later, Gracia-Bondia and Lazzarini 
\cite{GraciaLazzarini,GVF}
defined a Hopf algebra of Feynman
diagrams related to the Epstein-Glaser renormalization.
Then Pinter \cite{PinterHopf} derived the same algebra using partitions
of a set of points; her work was the starting point of
the present paper.
The Hopf algebra of \cite{GraciaLazzarini,PinterHopf}
looks similar to the Connes-Kreimer algebra, but it is actually
different because it allows for the renormalization of
nonirreducible diagrams, and it works in the configuration
space instead of the momentum space.
This paper is devoted to a generalization of Pinter's
 construction 
 to any bialgebra $\Bcal$.  In the first section, we consider the
 algebra $T(T(\Bcal)^+)$, where $T(\Bcal)^+$ is the nonunital
 subalgebra $\bigoplus_{n\geq 1}\tbk n$ of the tensor algebra
 $T(\Bcal)=\bigoplus_{n\geq 0}\tbk n$.  We describe how the
 coproduct of $\Bcal$ extends freely to define bialgebra structures on
 \tb\ and $T(T(\Bcal)^+)$.  These {\it free bialgebras} are
 noncommutative, and are cocommutative if and only if $\Bcal$ is
 cocommutative.  We then show that $T(T(\Bcal)^+)$ can be equipped
 with a very different coalgebra structure, making it a graded
 bialgebra which is neither commutative nor cocommutative, regardless
 of whether or not $\Bcal$ is cocommutative.  The abelianization of
 the bialgebra $T(T(\Bcal)^+)$ gives us the commutative bialgebra
 $S(T(\Bcal)^+)$, and $S(S(\Bcal)^+)$ is shown to be a subbialgebra of
 $S(T(\Bcal)^+)$. In quantum field applications, the bialgebra which
 is relevant to renormalization is $S(S(\Bcal)^+)$. 
When $\Bcal$ is the Hopf algebra of a commutative group, we define a
 homomorphism from $S(S(\Bcal)^+)$ onto the Fa\`a di Bruno bialgebra,
 which shows that $S(S(\Bcal)^+)$ is a kind of generalization of the
 algebra of formal diffeomorphisms.  Hopf algebras can be obtained from
 $T(T(\Bcal)^+)$ and $S(S(\Bcal)^+)$ as quotients by certain
biideals. 
We refer to these Hopf algebras as
the noncommutative and commutative Pinter algebra, respectively. 
All of our constructions are functorial in $\Bcal$ and the various
mappings between them correspond to natural transformations.
We describe the connection between
the commutative Pinter algebra and the Connes-Moscovici
algebra.
Finally, we prove that our construction gives the same
results as the standard renormalization procedure for
scalar fields. 
 
\section{The renormalization bialgebra}
In all that follows $\Bcal$ is a (not necessarily unital) bialgebra
over a field of characteristic zero,
with product \mub, coproduct $\deltaB$, and counit
$\counitB$. We denote the product of two elements $x$, $y$ of
$\Bcal$ by $x\cdot y$. We write $\tb^+$ for the subalgebra
$\bigoplus_{n\geq 1}\tbk n$ of the tensor algebra \tb, and 
denote the generators
$x_1\otimes\cdots\otimes x_n$ (where $x_i \in \Bcal$) of the vector space
\tbn\ by $(x_1,\dots,x_n)$; in particular, elements of $\tbk
1\cong\Bcal$ have the form $(x)$, for $x\in\Bcal$.  We use the symbol
$\potimes$, rather than $\otimes$, for the product in \tb,
so that
$(x_1,\dots,x_n)\potimes (x_{n+1},\dots,x_{n+m})=(x_1,\dots,x_{n+m})$.
By associativity, we may consider the product \mub\ as a map
$\tb^+\rightarrow\Bcal$; hence $\mub (x_1,\dots,x_n)=x_1\cdots x_n$,
for all $(x_1,\dots,x_n)\in\tbn$.
We denote the product operation in \ttb\ by
juxtaposition, so that $T^k(\tb^+)$ is generated by the elements
$a_1 a_2\cdots a_k$, where $a_i\in \tb^+$.  We denote the product of 
$a_1,\dots,a_k\in\tb^+$ in $T(T(\Bcal)^+)$ by $\Pi_{i=1}^k a_i$,
and we write $\bigotimes_{i=1}^k a_i$ for their product in $T(\Bcal)$.

\subsection{The free bialgebra structure on 
$T(T(\Bcal)^+)$}
The coproduct $\deltaB$ and counit $\counitB$ extend uniquely to
a coproduct and counit on \tb\ that are compatible with the multiplication
of \tb, thus making \tb\ a bialgebra. We remark that
this construction ignores completely the algebra structure of
$\Bcal$; the bialgebra \tb\ is in fact the free bialgebra
 on the underlying coalgebra of $\Bcal$.
Similarly, the coproduct and counit of the nonunital
bialgebra $\tb^+$ extend to define a free bialgebra structure
on $T(\tb^+)$.  We denote the coproduct of both \tb\ and $T(\tb^+)$
by $\delta$.  Hence, if we use the Sweedler notation $\deltaB(x)=
\sum x\ix1 \otimes x\ix2$ for the coproduct of $x$ in $\Bcal$
then, for $a=(x^1,\dots,x^n)\in\tbk n$ and
$u=a^1\cdots a^k\in T^k(\tb^+)$, we have
\[
\delta (a) = \sum (x\ix1^1,\dots, x\ix1^n)\otimes(x\ix2^1,\dots, x\ix2^n)
\]
and
\[
\delta (u) = \sum a\ix1^1\cdots a\ix1^k\otimes a\ix2^1\cdots a\ix2^k.
\]
The counit is defined by
$\counitT(a) = \counitB(x^1)\cdots\counitB(x^n)$
and
$\counitT(u) =  \counitT(a^1)\cdots\counitT(a^k)$.
A similar construction was put forward by Florent Hivert \cite{HivertPhD}.

For any linear map of vector spaces $f\colon V\rightarrow W$,
we denote by $T(f)$ the corresponding algebra map
$T(V)\rightarrow T(W)$, given by $(x_1,\dots,x_n)\mapsto
(f(x_1),\dots,f(x_n))$, for all $(x_1,\dots,x_n)\in T(V)$.
Note that, in particular, the map $T(\mub)\colon T(\tb^+)\rightarrow
\tb$ satisfies
\begin{align*}
T(\mub)&((x_1^1,\dots ,x_{r_1}^1)(x_1^2\,,\dots\, ,x_{r_2}^2)
\cdots (x_1^k,\dots, x_{r_k}^k))\\
&\,=\,(\mub(x_1^1,\dots ,x_{r_1}^1),\mub(x_1^2\,,\dots\, ,x_{r_2}^2),
\dots ,\mub(x_1^k,\dots, x_{r_k}^k))\\
&\,=\,(x_1^1\cdots x_{r_1}^1,\, x_1^2\cdots  x_{r_2}^2,
\dots,\, x_1^k\cdots x_{r_k}^k).
\end{align*}

\subsection{Grading $T(T(\Bcal)^+)$ by compositions}
 
A {\it composition\/} $\rho$ is a (possibly empty) finite sequence of
positive integers, usually referred to as the {\it parts\/} of $\rho$.
We denote by $\ell (\rho)$ the length, that is the number of parts, of
$\rho$, write $|\rho|$ for the sum of the parts, and say
that $\rho$ is a {\it composition of\/} $n$ in the case that
$|\rho|=n$.  We denote by $C_n$ the set of all compositions of $n$,
and by $C$ the set $\bigcup_{n\geq 0}C_n$ of all compositions of all
nonnegative integers.  
For example, $\rho=(1,3,1,2)$ is a composition of 
$7$ having length $4$.
The first four $C_n$ are
$C_0=\{e\}$, where $e$ is the empty composition,
$C_1=\{(1)\}$, $C_2=\{(1,1),(2)\}$ and
$C_3=\{(1,1,1),(1,2),(2,1),(3)\}$. 
The total number of compositions of $n$ is $2^{n-1}$,
the number of compositions
of $n$ of length $k$ is $\frac{(n-1)!}{(k-1)!(n-k)!}$
and the number of compositions of $n$ with precisely $\alpha_i$ occurrences
of the integer $i$, for all $i$ (and hence $\sum i\alpha_i
=n$) is
$\frac{(\alpha_1+\dots+\alpha_n)!}{\alpha_1!\dots\alpha_n!}$.

The set $C$ is a monoid under the operation $\potimes$
of concatenation of sequences:
$(r_1,\dots,r_n)\potimes (r_{n+1},\dots,r_{n+m})=(r_1,\dots,r_{n+m})$.
The identity element of $C$ is the empty composition $e$.

For any composition $\sigma = (s_1,\dots,s_k)$, and $1\leq i\leq k$,
let $\sigma_i$ be the interval $\{s_1+\cdots +s_{i-1}+1,\cdots,
s_1+\cdots +s_i\}$.  The set of all compositions is partially ordered
by setting $\rho\leq\sigma$ if and only if each part of $\sigma$ is a
sum of parts of $\rho$, that is, if and only if each interval
$\sigma_i$ is a union of $\rho_j$'s.  This order relation is called
$\emph{refinement}$, and will play an essential role in the definition
of the bialgebra of renormalization.  For compositions $\rho\leq
\sigma$, with $\sigma =(s_1,\dots,s_\ell)$ and $\rho
=(r_1,\dots,r_k)$, and $1\leq i\leq\ell$, we define the {\it
  restriction\/} $\rho|\sigma_i$ as the composition
$(r_{j_i},\dots,r_{k_i})$ of $s_i$, where $j_i=\min
\{j\,\colon\,\rho_j\subseteq\sigma_i\}$ and $k_i=\max
\{j\,\colon\,\rho_j\subseteq\sigma_i\}$.  
Note that we then have the factorization
$\rho=(\rho|\sigma_1)\potimes\cdots\potimes(\rho|\sigma_k)$.

For example, if $\sigma=(4,5)$ and $\rho=(1,2,1,2,2,1)$,
then $\rho=(\rho|\sigma_1)\potimes(\rho|\sigma_2)$
where $(\rho|\sigma_1)=(1,2,1)$ is a composition of 4
and $(\rho|\sigma_2)=(2,2,1)$ is a composition of 5.
Thus $\rho\le\sigma$ and we say that $\rho$ is a refinement
of $\sigma$. Note that $|\rho|=|\sigma|$ and $\ell(\rho)\ge\ell(\sigma)$
if $\rho\le\sigma$.

If $\rho\le\sigma$, we define the {\it quotient\/} $\sigma/\rho$ to be 
the composition of $\ell (\rho)$ given by $(t_1,\dots,t_k)$, where 
$t_i= \ell ((\rho|\sigma_i))$, for $1\leq i\leq k$.
In our example $\sigma/\rho=(3,3)$. 
Note that, for $\sigma\in C_n$ with $\ell(\sigma)=k$, we have
$(n)/\sigma=(k)$, $\sigma/(1,1,\dots,1)=\sigma$, and
$\sigma/\sigma= (1,1,\dots,1)\in C_k$ 

Each of the sets $C_n$ (as well as all of $C$) is partially ordered by
refinement.  
Each $C_n$ has unique
minimal element $(1,\dots , 1)$ and unique maximal element $(n)$, and
these are all the minimal and maximal elements in $C$.  The partially
ordered sets $C_n$ are actually Boolean algebras, but we will not use
this fact here.

Now we give a lemma that will be used in proving the coassociativity
of the coproduct.
 
\begin{lem}\label{lem1}
If $\rho\leq\tau$ in $C$, then the map
$\sigma\mapsto\sigma/\rho$ is a bijection from the set
$\{ \sigma\,\colon\,\rho\leq\sigma\leq\tau\}$ onto the
set $\{ \gamma\,\colon\,\gamma\leq\tau/\rho\}$.
\end{lem} 
\begin{proof}
Suppose that $\rho =(r_1,\dots,r_k)$ and
that $\gamma = (s_1,\dots,s_\ell)\leq\tau/\rho$. Define
$\bar\gamma\in C$ by
\[
\bar\gamma \,=\,(r_1+\cdots+r_{s_1},\,r_{s_1+1}+\cdots+ r_{s_1+s_2},
\,\dots\, ,\,r_{k-s_\ell +1}+\cdots +r_k).
\]
It is then readily verified that $\rho\leq\bar\gamma\leq\tau$,
and that the map $\gamma\mapsto\bar\gamma$ is inverse to the
map $\sigma\mapsto\sigma/\rho$.
\end{proof}

The monoid of compositions allows us to define a grading  on
$T(T(\Bcal)^+)$. 
 
For all $n\geq 0$ and $\rho = (r_1,\dots ,r_k)$ in $C_n$,
we let \tbk\rho\ denote the subspace of $T^k(\tb^+)$
given by $\tbk{r_1}\otimes\cdots\otimes\tbk{r_k}$. 
We then have the direct sum decomposition
$\ttb = \bigoplus_{\rho\in C}\tbk\rho$, where
$\tbk\rho\cdot\tbk\tau\subseteq\tbk{\rho\potimes\tau}$
for all $\rho, \tau\in C$, and $1\sub{T(T(\mathcal B))}\in\tbk e$,
where $e$ denotes the empty composition;
in other words, \ttb\ is a $C$-graded algebra.

We use this grading to define operations on  \ttb.
Given $a=(x_1,\dots ,x_n)\in\tbn$ and $\rho = (r_1,\dots ,r_k)\in
C$, we define $a|\rho_i\in\tbk{r_i}$, for $1\leq i\leq k$ by
\[
a|\rho_i\,=\, \bigotimes_{j\in\rho_i}(x_j)
\,=\,(x_{r_1+\cdots+r_{i-1}+1},\dots,x_{r_1+\cdots+r_i}), 
\]
where we take $x_j=0$, for $j>n$, and
we define the \emph{restriction}
$a|\rho\in\tbk\rho$ and 
\emph{contraction} $a/\rho\in\tbk k$ by
\begin{align*}
a|\rho &\,=\, (a|\rho_1)\cdots (a|\rho_k)\\
&\,=\, (x_1,\dots ,x_{r_1})(x_{r_1+1}\,,\dots\, ,x_{r_1+r_2})
\cdots (x_{n-r_k+1},\dots, x_n)
\end{align*}
and
\begin{align*}
a/\rho &\,=\, T(\mub)(a|\rho)
\,=\, (\mub (a|\rho_1),\dots,\mub (a|\rho_k)) \\
&\,=\, (x_1\cdots x_{r_1},\, 
x_{r_1+1}\cdots  x_{r_1+r_2},
\dots,\, x_{n-r_k+1}\cdots x_n),
\end{align*}
where $x_i \cdots x_j$ denotes the product 
of $x_i,\dots,x_j$ in $\Bcal$.
Note that $a|\rho$ and $a/\rho$ are zero if $|\rho|\neq n$.
Observe also that, even though the quantity $a|\rho_i$ depends (up
to a scalar multiple) on the choice of elements $x_1,\dots, x_n\in
\Bcal$ representing $a=(x_1,\dots,x_n)\in\tbn$, the quantities
$a|\rho$ and $a/\rho$ depend only on $a$ and $\rho$.  

More generally, for $u=a_1\cdots a_\ell$ in \tbk\sigma, and
$\rho\in C$, 
we define $u|\rho\in\tbk\rho$ and $u/\rho\in\tbk{\sigma/\rho}$ by
\[
u|\rho \,=\, a_1|(\rho|\sigma_1)\cdots a_\ell |(\rho|\sigma_\ell)
\quad
\mbox{and}
\quad
u/\rho\,=\, a_1/(\rho|\sigma_1)\cdots a_\ell /(\rho|\sigma_\ell).
\]
Note that $u|\rho$ and $u/\rho$ are both zero if $\rho\nleq\sigma$.

\begin{exa}{\rm
Suppose that $\rho = (1,2,1,2,2,1)$, 
$\sigma= (3,1,2,3)$ and $\tau= (4,5)$, so that
 $\rho\leq\sigma\leq\tau$ in $C$.  We then have the quotients 
$\sigma/\rho=(2,1,1,2)$, $\tau/\rho=(3,3)$ and $\tau/\sigma = (2,2)$.
If $u=(x_1, x_2, x_3, x_4)(y_1, y_2, y_3, y_4, y_5)$ in \tbk\tau,
then
\begin{align*}
u|\rho &\,=\, (x_1)(x_2, x_3)(x_4)(y_1, y_2)(y_3, y_4)(y_5)\in\tbk\rho,\\
u|\sigma &\,=\, (x_1, x_2, x_3)(x_4)(y_1, y_2)(y_3, y_4, y_5)\in\tbk\sigma,\\
u/\rho &\,=\, (x_1, x_2\cdot x_3, x_4)(y_1\cdot y_2, y_3\cdot y_4,
y_5)\in\tbk{\tau/\rho},\\
u/\sigma &\,=\, (x_1\cdot x_2\cdot x_3, x_4)(y_1\cdot y_2, y_3\cdot y_4\cdot
y_5)\in\tbk{\tau/\sigma},\\
(u|\sigma)/\rho &\,=\, (u/\rho)|(\sigma/\rho) \,=\,
(x_1, x_2\cdot x_3)(x_4)
(y_1\cdot y_2)(y_3\cdot y_4, y_5)\in\tbk{\sigma/\rho}.
\end{align*}
}
\end{exa}
The last equality illustrates the following lemma.
\begin{lem}\label{lem2}
For all $\rho,\sigma,\tau\in C$, and $u\in\tbk\tau$,
the equalities
\[
(u|\sigma)|\rho\,=\, u|\rho, \quad
(u/\rho)/(\sigma/\rho)\,=\, u/\sigma, \quad\mbox{and}\quad
(u|\sigma)/\rho\,=\, (u/\rho)|(\sigma/\rho)
\]
hold in \ttb.
\end{lem}
\begin{proof}
First note that all the expressions above are
zero unless $\rho\leq\sigma\leq\tau$ in $C$. Thus 
it suffices to prove the result for $u=a=(x_1,\dots,x_n)
\in\tbk n$ and $\rho\leq\sigma$ in $C_n$.
If $\ell (\sigma)=k$, then
$a|\sigma=\prod_{i=1}^k a|\sigma_i$, and 
$$
(a|\sigma)|\rho\,=\,
\prod_{i=1}^k \,(a|\sigma_i)|(\rho|\sigma_i)
\,=\, \prod_{i=1}^k\prod_{\rho_j\subseteq\sigma_i}\, (a|\sigma_i)|\rho_j
\,=\, \prod_{i=1}^k\prod_{\rho_j\subseteq\sigma_i}\, (a|\rho_j),
$$
which is equal to $a|\rho$. For the second equation, we have
\begin{align*}
  (a/\rho)/(\sigma/\rho)&\,=\, (\mub ((a/\rho)|(\sigma/\rho)_1),\dots,
\mub ((a/\rho)|(\sigma/\rho)_k))\\
&\,=\, (\mub (a|\sigma_1),\dots,\mub (a|\sigma_k))\\
&\,=\, a/\sigma. 
\end{align*}
Finally, we have
\begin{align*}
(a|\sigma)/\rho &\,=\, \prod_{i=1}^k (a|\sigma_i)/(\rho|\sigma_i)\\
&\,=\, \prod_{i=1}^k\bigotimes_{\rho_j\subseteq\sigma_i}
(\mub ((a|\sigma_i)|\rho_j))\\
&\,=\, \prod_{i=1}^k\bigotimes_{\rho_j\subseteq\sigma_i}
(\mub (a|\rho_j))\\  
&=\prod_{i=1}^k (a/\rho)|(\sigma/\rho)_i,
\end{align*}
which is equal to $(a/\rho)|(\sigma/\rho)$.
\end{proof}
\begin{lem}\label{lem3}
For all $\rho\in C$ the restriction and contraction maps,
given by $u\mapsto u|\rho$ and $u\mapsto u/\rho$, respectively,
for homogeneous $u$, are coalgebra maps $\ttb\rightarrow\ttb$; that is,
for all $\rho, \sigma\in C$ and $u\in\tbk\sigma$,
with free coproduct $\delta (u) = \sum u\ix1\otimes u\ix2$, the equalities
\begin{align*}
\delta
(u|\rho) &\,=\, \sum u\ix1|\rho \otimes u\ix2|\rho,\;\;\mbox{and}\\\\
\delta
(u/\rho) &\,=\, \sum u\ix1/\rho \otimes u\ix2/\rho 
\end{align*}
hold.
\end{lem}
\begin{proof}
First note that both sides of each of the above equations
are zero if it is not the case that $\rho\leq\sigma$ in $C$. 
Hence, by multiplicativity of $\delta$ it suffices to consider the 
case in which
$u=a=(x^1,\dots ,x^n)\in\tbk n$ and $\rho=(r_1,\dots,r_k)\in C_n$.
Then we have
\begin{eqnarray*}
  \delta (a|\rho)\!\! &=&\!
\delta ((x^1,\dots, x^{r_1})(x^{r_1+1},\dots, x^{r_2})
\,\cdots\, (x^{n-r_k+1},\dots, x^n))\\
&=&\! \sum ((x\ix1^1,\dots, x\ix1^{r_1})(x\ix1^{r_1+1},\dots, x\ix1^{r_2})
\,\cdots\, (x\ix1^{n-r_k+1},\dots, x\ix1^n))\\&&\otimes\,
((x\ix2^1,\dots, x\ix2^{r_1})(x\ix2^{r_1+1},\dots, x\ix2^{r_2})
\,\cdots\, (x\ix2^{n-r_k+1},\dots, x\ix2^n))\\
&=&\!\sum (x\ix1^1,\dots,x\ix1^n)|\rho \otimes (x\ix2^1,\dots,x\ix2^n)|\rho\\
&=&\!\sum a\ix1|\rho \otimes a\ix2|\rho,
\end{eqnarray*}
and
\begin{eqnarray*}
  \delta (a/\rho)\!\! &=&\!
\delta (x^1\cdots x^{r_1}, 
\,\dots\,, x^{n-r_k+1}\cdots x^n)\\
&=&
\sum ((x^1\cdots x^{r_1})\ix1,
\,\dots\,, (x^{n-r_k+1}\cdots x^n)\ix1)\\&&\otimes\,
((x^1\cdots x^{r_1})\ix2,
\,\dots\,, (x^{n-r_k+1}\cdots x^n)\ix2)\\
&=&\! \sum (x\ix1^1\cdots x\ix1^{r_1},
\,\dots\,, x\ix1^{n-r_k+1}\cdots x\ix1^n)\\
&&\otimes\, (x\ix2^1 \cdots x\ix2^{r_1},
\,\dots\,, x\ix2^{n-
r_k+1}\cdots x\ix2^n)\\
&=&\!\sum (x\ix1^1,\dots,x\ix1^n)/\rho \otimes (x\ix2^1,\dots,x\ix2^n)/\rho\\
&=&\!\sum a\ix1/\rho \otimes a\ix2/\rho.
\end{eqnarray*}
\end{proof}

\subsection{The renormalization coproduct and counit}
 
We now define the coproduct $\Delta$ on the  algebra \ttb\ 
by
\begin{equation}\label{copdef}
\Delta u\,\,=\, \sum_{\sigma\leq\tau}u\ix1|\sigma \otimes u\ix2/\sigma,
\end{equation}
for $u\in\tbk\tau$, with free coproduct $\delta (u)=\sum u\ix1\otimes
u\ix2$.
The coproduct $\Delta$ is called the \emph{renormalization coproduct}
because of its role in the renormalization of quantum field theories.
Note that $\Delta$ is an algebra
map, and hence is determined by
\[
\Delta a\, \,=\, \sum_{\sigma\in C_n}a\ix1 |\sigma \otimes a\ix2 /\sigma,
\]
for all $a\in\tbk n$ with $n\geq 1$. For example
\begin{align*}
\Delta(x) &\,=\, \sum (x\ix1) \otimes (x\ix2),\\
\Delta(x,y) &\,=\, \sum (x\ix1)(y\ix1) \otimes (x\ix2,y\ix2)
  + \sum (x\ix1,y\ix1) \otimes (x\ix2\cdot y\ix2),
\end{align*}
\begin{eqnarray*}
\Delta(x,y,z)\!\! &=&\!\! \sum (x\ix1)(y\ix1)(z\ix1) \otimes
(x\ix2,y\ix2,z\ix2)
\\&&
+\, \sum (x\ix1)(y\ix1,z\ix1) \otimes (x\ix2,y\ix2\cdot z\ix2)
\\&&
+\, \sum (x\ix1,y\ix1)(z\ix1) \otimes (x\ix2\cdot y\ix2,z\ix2)
\\&&
+\, \sum (x\ix1,y\ix1,z\ix1) \otimes (x\ix2\cdot y\ix2\cdot z\ix2).
\end{eqnarray*}
 
The counit $\counit$ of \ttb\ is the algebra map $\ttb\rightarrow\bC$
whose restriction to $\tb^+$ is given by $\counit
((x))=\counitB(x)$, for $x\in\Bcal$, and
$\counit((x_1,\dots,x_n))=0$, for $n > 1$.
\begin{thm}\label{theorem1}
The algebra \ttb, together with the structure maps
$\Delta$ and $\counit$ defined above, is a bialgebra,
called the \emph{renormalization bialgebra}.
\end{thm} 
\begin{proof}
For $u\in\tbk\tau$, we have
\begin{align}
(\Delta\otimes\Id)\Delta u &\,=\,
\sum_{\sigma\leq\tau}\Delta (u\ix1 |\sigma )\otimes u\ix2/\sigma \nonumber\\
&\,=\, \sum_{\rho\leq\sigma\leq\tau}
(u\ix1|\sigma)|\rho \otimes (u\ix2|\sigma)/\rho \otimes u\ix3/\sigma \nonumber \\
&\,=\, \sum_{\rho\leq\sigma\leq\tau}
u\ix1|\rho \otimes (u\ix2|\sigma)/\rho \otimes u\ix3/\sigma ,
\label{cop1}
\end{align}
where the second equality is by Lemma \ref{lem3}, and the third
by Lemma \ref{lem2}.  On the other hand,
\begin{align}
(\Id\otimes\Delta)\Delta u &\,=\,
\sum_{\rho\leq\tau} u\ix1 |\rho \otimes \Delta(u\ix2/\rho) \nonumber\\
&\,=\, \sum_{\rho\leq\tau}\sum_{\gamma\leq\tau/\rho}
u\ix1|\rho \otimes (u\ix2/\rho)|\gamma \otimes (u\ix3/\rho)/\gamma\nonumber \\
&\,=\, \sum_{\rho\leq\sigma\leq\tau}
u\ix1|\rho \otimes (u\ix2/\rho)|(\sigma/\rho) \otimes (u\ix3/\rho)/(\sigma/\rho) ,
\label{cop2}
\end{align}
where the second equality is by Lemma \ref{lem3} and the third follows from
Lemma \ref{lem1}.  Expressions \ref{cop1} and \ref{cop2} are equal
by Lemma \ref{lem2}, and hence $\Delta$ is coassociative.  
 
For $a=(x^1,\dots,x^n)\in\tbk n$, with $\delta (a)
=\sum a\ix1 \otimes a\ix2$, we have
\[
(\Id\otimes\counit)\Delta a \,=\,
\sum_{\sigma\in C_n}a\ix1|\sigma \,\counit (a\ix2/\sigma).
\]
Now $\counit (a\ix2/\sigma)=0$ unless $\sigma = (n)$; hence
\begin{align*}
(\Id\otimes\counit)\Delta a &\,=\,
\sum (x\ix1^1,\dots,x\ix1^n)\varepsilon_{\scriptscriptstyle T(\mathcal B)}
(x\ix2^1,\ldots,x\ix2^n)\\
&\,=\,
\sum (x\ix1^1,\dots,x\ix1^n)\counitB(x\ix2^1)\cdots 
\counitB(x\ix2^n)\\
&\,=\, \sum (x\ix1^1\counitB(x\ix2^1)
,\dots,x\ix1^n\counitB(x\ix2^n))\\
&\,=\, a.
\end{align*}
The proof that $(\counit\otimes\Id)\Delta a =a$ is similar.
We have already observed that $\Delta$ and $\epsilon$ are
algebra maps; hence \ttb\ is a bialgebra.
\end{proof}

We now have two coproducts on \ttb, namely the free coproduct $\delta$
and the renormalization coproduct $\Delta$ defined by Equation
\ref{copdef}.  In order to avoid confusion in the following sections,
we adopt the following alternate Sweedler notation for the new
coproduct:
\[
\Delta u\,=\,\sum u\ii1\otimes u\ii2,
\]
for all $u\in\ttb$.  Note that, in particular, 
\[
\Delta (x) \,=\, \sum (x)\ii1\otimes (x)\ii2 \,=\, \sum (x\ix1)\otimes
(x\ix2),
\]
for all $x\in\Bcal$.  Whenever we simply refer to the bialgebra
\ttb, we shall mean the renormalization bialgebra; we will always
state explicitly when considering \ttb\ with the free bialgebra
structure.

\subsection{Recursive definition of the coproduct}

The action of $\Bcal$ on itself by left
multiplication extends to an action 
$\Bcal\otimes\tb\rightarrow\tb$, denoted by $x\otimes a\mapsto x\act a$,
in the usual manner, that is
\[
x\act a \,=\, (x\cdot x_1,x_2,\dots,x_n),
\]
for all $x\in\Bcal$ and $a=(x_1,\dots,x_n)\in\tbn$.
This action, in turn, extends to an action of $\Bcal$ on \ttb,
denoted similarly by $x\otimes u\mapsto x\act u$; that is:
\[
x\act u \,=\, (x\act a_1)a_2\cdots a_k,
\]
for all $x\in\Bcal$ and $u=a_1\cdots a_k\in\ttb$.

The following proposition, together with the fact that 
$\Delta 1 = 1\otimes 1$ and $\Delta (x) = \sum (x\ix1)\otimes (x\ix2)$
for all $x\in\Bcal$, determines $\Delta$ recursively on
$T(\Bcal)$ and hence, by multiplicativity, determines $\Delta$
on all of \ttb.

\begin{pro}\label{proposition1}
For all $a\in\tbn$, with $n\geq 1$, and $x\in\Bcal$,
\begin{equation}
\Delta ((x)\potimes a) \,=\, \sum (x\ix1)a\ii1 \otimes (x\ix2)\potimes a\ii2
+\sum (x\ix1)\potimes a\ii1 \otimes x\ix2 \act a\ii2.
\label{recur}
\end{equation}
\end{pro}
\begin{proof}
We denote by $C'_n$ the set of all
compositions of $n$ whose first part is equal to $1$ and write
$C''_n$ for the set difference $C_n\backslash C'_n$.  Note that the
map $\rho\mapsto (1)\potimes\rho$ is a bijection from
$C_n$ onto $C'_{n+1}$.  If $\rho=(r_1,r_2,\dots,r_k)$
we define $\rho^+=(r_1+1,r_2,\dots,r_k)$ and
the map $\rho\mapsto \rho^+$ is a bijection from
$C_n$ onto $C''_{n+1}$. From the definition of $a|\rho$ and
$a/\rho$ it can be checked that
$\big((x)\circ a\big)/\big((1)\circ\rho\big)=(x)\circ(a/\rho)$,
$\big((x)\circ a\big)|\big((1)\circ\rho\big)=(x)(a|\rho)$,
$\big((x)\circ a\big)/\rho^+=(x)\action(a|\rho)$
and 
$\big((x)\circ a\big)|\rho^+=(x)\circ(a|\rho)$,
where in the last identity we extend the 
product of $T(\Bcal)$ by
$(x)\circ u=\big((x)\circ a_1\big)a_2\dots a_k$
if $u=a_1 a_2\dots a_k$.

We then have
\begin{eqnarray*}
\Delta ((x)\potimes a)\! &=&\! 
\sum_{\rho\in C'_{n+1}}((x)\potimes a)\ix1 |\rho\otimes
((x)\potimes a)\ix2 /\rho\\
&&+\!
\sum_{\rho\in C''_{n+1}}((x)\potimes a)\ix1 |\rho\otimes
((x)\potimes a)\ix2 /\rho\\
&=&\!
\sum_{\sigma\in C_n} (x\ix1)(a\ix1|\sigma)\otimes
(x\ix2)\potimes (a\ix2/\sigma)\\
&&+
\sum_{\tau\in C_n} (x\ix1)\potimes (a\ix1|\tau)\otimes
(x\ix2)\act (a\ix2/\tau),
\end{eqnarray*}
which is precisely the right-hand side of Equation \ref{recur}.
\end{proof}

The recursive definition of $\Delta$ was used in
\cite{BFII} to show that \ttb\ is isomorphic to the noncommutative
Hopf algebra of formal diffeomorphisms in the case that $\Bcal$ is
the trivial algebra.

We may formulate Equation \ref{recur} as follows: 
Corresponding to an element $x$ of $\Bcal$ there are three
linear operators on \ttb:
\begin{align*}
 A_x (u) &\,=\, x\act u \\
  B_x (u) &\,=\, ((x)\circ a_1)a_2\cdots a_k \qquad 
\text{(where $u=a_1\cdots a_k$)}\\
  C_x (u) &\,=\, (x)u 
\end{align*}
induced by left multiplication in $\Bcal$, \tb, and \ttb,
respectively.
With this notation, Equation \ref{recur} takes the form
\begin{equation}
\Delta (B_x (a)) \,=\, \sum (C_{x\ix1}\otimes B_{x\ix2} +
B_{x\ix1}\otimes A_{x\ix2})
\Delta a.
\end{equation}

\noindent
As a third formulation, let $A$, $B$ and $C$ be the mappings 
from \Bcal\ to the set of linear
operators on \ttb\ respectively given by $x\mapsto A_x$, $x\mapsto B
_x$, and $x\mapsto C_x$; then Equation \ref{recur} takes the form
$$
\Delta (B_x (a)) \,=\, (B\otimes A+C\otimes B)(\delta (x))(\Delta a).
$$
We also have
\begin{align*}
\Delta (A_x (a)) &\,=\, (A\otimes A)(\delta (x))(\Delta a),\\
\Delta (C_x (a)) &\,=\,  (C\otimes C)(\delta (x))(\Delta a).
\end{align*}

Finally, we give a more explicit expression for the
coproduct. If $a=(x_1,\dots,x_n)$, we have
\begin{equation}
\Delta a \,=\, \sum_u 
a\ix1^1\cdots a\ix1^{\ell(u)}
\otimes (\prod a\ix2^1,\dots,\prod a\ix2^{\ell(u)}),
\label{composition}
\end{equation}
where the product $a\ix1^1\cdots a\ix1^{\ell(u)}$
is in \ttb.
In this formula, which is a simple rewriting of 
Equation \ref{copdef}, $u$ runs over the compositions
of $a$. 
By a composition of $a$, we mean an element $u$ of \ttb\
such that $u=(a|\rho)$ for some $\rho\in C_n$.
If the length of $\rho$ is $k$, we 
can write $u=a^1\dots a^k$
where $a^i\in\tb$ are called the blocks of $u$.
Finally the length of $u$ is
$\ell(u)=\ell(\rho)=k$.
To complete the definition of Equation \ref{composition},
we still have to define $a\ix1^i$ and $\prod a\ix2^i$.
If $a^i=(y^1,\dots,y^m)$ is a block,
then $a\ix1^i=(y\ix1^1,\dots,y\ix1^m)\in \tb^+$
and $\prod a\ix2^i=\mub (y\ix2^1,\dots ,y\ix2^m) \in \Bcal$.

\subsection{Functoriality}
Given vector spaces $V$, $W$, and a linear map $f\colon
V\rightarrow W$, we denote by $\hat f$ the algebra map
$T(T(f))\colon T(T(V)^+)\rightarrow T(T(W)^+)$, 
determined by $\hat f (a)=
T(f)(a)= (f(x^1),\dots,f(x^n))$ for all $a=(x^1,\dots,x^n)\in T^n
(V)$, and $\hat f(u)=\hat f(a_1)\cdots\hat f(a_k)$, for all
$u=a_1\cdots a_k\in T(T(V)^+)$, where $a_1,\dots,a_k\in T(V)^+$.
The following proposition shows that the renormalization construction
on bialgebras is functorial.
\begin{pro}
  If $\Bcal$ and $\Ccal$ are bialgebras and
$f\colon\Bcal\rightarrow\Ccal$ is a bialgebra map, then
$\hat f\colon\ttb\rightarrow\ttc$ is a bialgebra
map. 
\end{pro}
\begin{proof}
Given $a\in T^n(\Bcal)$, and a composition $\rho\in C_n$, it follows
directly from the definition of $\hat f$ that $\hat f(a|\rho)=
\hat f(a)|\rho$, and using the fact that $f$ preserves products, it follows
that $\hat f(a/\rho)=\hat f(a)/\rho$. By multiplicativity of $\hat f$ 
we thus have
\[
\hat f (u|\sigma)\,=\, \hat f(u)|\sigma\qquad\text{and}\qquad
\hat f (u/\sigma)\,=\, \hat f(u)/\sigma,
\]
for all homogeneous $u\in\ttb$ and all compositions $\sigma\in C$.
Furthermore, it is immediate from the fact $f\colon\Bcal\rightarrow\Ccal$ 
preserves coproducts that $\hat f\colon\ttb\rightarrow\ttc$ preserves
free coproducts.  Thus for all compositions $\tau$, and $u\in
T^\tau (\Bcal)$, with free coproduct  $\delta (u)=\sum u\ix1\otimes u\ix2$,
we have
\begin{align*}
  \Delta\hat f(u) 
&\,=\, \sum_{\sigma\leq\tau}f(u\ix1)|\sigma \otimes f(u\ix2)/\sigma,\\
&\,=\,\sum_{\sigma\leq\tau}f(u\ix1|\sigma) \otimes f(u\ix2/\sigma),\\
&\,=\,(\hat f\otimes\hat f)\Delta (u),
\end{align*}
and so $\hat f$ preserves the renormalization coproduct.
\end{proof}
\subsection{Grading}
We now assume that $\Bcal$ is a graded bialgebra; this entails no loss
of generality because we can always consider that all elements of
$\Bcal$ have degree 0. The grading of $\Bcal$ will be used to define a
grading on the bialgebra \ttb.  We denote by $|x|$ the degree
of a homogeneous element $x$ of \Bcal, and by $\grad (a)$ the degree
(to be defined) of homogeneous $a$ in \ttb.
We first discuss the grading of elements of \tb.  The degree of 1 is
zero, the degree of $(x)\in T^1(\Bcal)$ is equal to the degree of $x$
in $\Bcal$, that is, $\grad\big((x)\big)=|x|$. More generally, the degree of
$(x_1,\dots,x_n)\in T^n(\Bcal)$ is
$$
\grad\big((x_1,\dots,x_n)\big)
\,=\, |x_1|+\cdots+|x_n|+n-1.
$$
Finally, if $a_1,\dots,a_k$ are homogeneous elements
of $\tb^+$, the degree of their product in \ttb\ is 
defined by
$$
\grad(a_1\dots a_k) \,=\, \grad(a_1)+\cdots+\grad(a_k).
$$
\begin{pro}
  The renormalization bialgebra \ttb, with degree defined
as above, is a graded bialgebra.
\end{pro}
\begin{proof}
By definition, the degree is compatible with the
multiplication of \ttb. The fact that it is also
compatible with the renormalization coproduct of \ttb,
follows directly from Formula \ref{composition}.
\end{proof}
For dealing with fermions, we must use a $\mathbb{Z}_2$-graded
algebra $\Bcal$. In this case, we extend the
grading of \Bcal\ to a $\mathbb{Z}_2$-grading of
\ttb\ as follows:  for $a=(x^1,\dots,x^n)\in\tbn$, we set
$|a|=|x^1|+\dots+|x^n|$, and for $u=a_1\cdots a_k\in\ttb$,
we set $|u|=|a_1|+\cdots+|a_k|$.  The coproduct is determined by
\[
\Delta a \,=\, \sum_{\rho\in C_n}\sgn (a\ix1,a\ix2)\,a\ix1|
\
\rho\otimes a\ix2/\rho,
\]
for $a=(x^1,\dots,x^n)\in\tbn$, where 
$\delta (a) = \sum a\ix1\otimes a\ix2$ is the free coproduct and
\[
\sgn (a\ix1,a\ix2)\,=\, (-1)^{\sum_{k=2}^n\sum_{l=1}^{k-1} |x\ix1^k||x\ix2^l|}.
\]
is the usual Koszul sign factor.
 
\section{The commutative renormalization bialgebra}
\label{comrensect}
When the bialgebra $\Bcal$ is commutative, it is possible to work with
the symmetric algebra \ssb\ instead of the tensor algebra
$T(T(\Bcal)^+)$.  We construct \ssb\ as a subbialgebra of the quotient
bialgebra \stb\ of \ttb.  We denote by $\Sigma_n$ the set of all
permutations of $\{1,\dots,n\}$ and, for $a=(x_1,\dots,x_n)\in\tbn$
and $\sigma\in\Sigma_n$, we write $a^\sigma$ for $(x_{\sigma
  (1)},\dots, x_{\sigma (n)})$.  Let
$\alpha=\alpha_{\mathcal B}\colon\tb\rightarrow\tb$ 
be the symmetrizing operator given
by $a\mapsto \sum a^\sigma$, for all $a\in\tbn$, where the sum is over
all $\sigma\in\Sigma_n$.  We identify $\sbcal^+$, as a vector space,
with the image $\alpha (\tb^+)$
in $\tb$.  We write $\{a\}$ for $\alpha (a)$ and denote the
product in $\sbcal^+$ by $\pvee$, so that
\begin{equation}
\{x_1,\dots,x_n\} \,=\, \sum_{\sigma\in\Sigma_n} 
(x_{\sigma(1)},\dots,x_{\sigma(n)}),
\label{symgroup}
\end{equation}
and
$\{x_1,\dots,x_n\} \pvee \{x_{n+1},\dots,x_{n+m}\} =
\{x_1,\dots,x_{n+m}\}$, for all $x_1,\dots,x_{n+m}\in\Bcal$.  Note that
$\sbcal^+$, with product $\pvee$ is not a subalgebra of $\tb^+$.
  
We define \stb\ as the quotient algebra $\ttb/I$, where $I$ is the ideal
$\{u(ab-ba)v\,|\, \text{$u,v\in\ttb$ and $a,b \in\tb^+$}\}$.  The
symmetric algebra \ssb\ is the image of the subspace
$\tsb\subseteq\ttb$ under the canonical projection
$\ttb\rightarrow S(\tb^+)$ or, equivalently, the image of the
map $S(\alpha)\colon\stb\rightarrow\stb$, determined by
$S(\alpha)(a_1\cdots a_k) = \{a_1\}\cdots\{a_n\}$, for all
$a_1,\ldots, a_k\in\tb^+$.

The following lemma shows that the symmetric algebra $S(T(\Bcal)^+)$
inherits the renormalization coproduct.

\begin{lem}
  The symmetric algebra $S(T(\Bcal)^+)$ is a quotient bialgebra
of the renormalization bialgebra \ttb.
\end{lem}
\begin{proof}
 The bialgebra $S(T(\Bcal)^+)$ is obtained 
from the bialgebra  $T(T(\Bcal)^+)$ by the standard
quotient method (see, e.g., \cite{Kassel}, p.56).
We define the ideal 
$I=\{u(ab-ba)v\,|\, u,v\in T(T(\Bcal)^+), a,b \in T(\Bcal)^+\}$,
and note that
\begin{eqnarray*}
\Delta u(ab-ba)v\! &=&\!
\sum u\ii1 a\ii1 b\ii1 v\ii1 \otimes  u\ii2 a\ii2 b\ii2 v\ii2
\\&&
-
\sum u\ii1 b\ii1 a\ii1 v\ii1 \otimes  u\ii2 b\ii2 a\ii2 v\ii2
\\&=&\!
\sum u\ii1 (a\ii1 b\ii1-b\ii1 a\ii1) v\ii1 \otimes  u\ii2 a\ii2 b\ii2 v\ii2
\\&&
+
\sum u\ii1 b\ii1 a\ii1 v\ii1 \otimes  u\ii2 (a\ii2 b\ii2-b\ii2 a\ii2)v\ii2.
\end{eqnarray*}
Thus $\Delta I\subset I\otimes T(T(\Bcal)^+) +  T(T(\Bcal)^+)\otimes I$.
Moreover, $\counit(I)=0$ because
$\counit(ab-ba)=0$. Therefore $I$ is a coideal. Since $I$ is
also an ideal, the quotient $S(T(\Bcal)^+)=T(T(\Bcal)^+)/I$ is
a bialgebra, which is commutative \cite{Kassel}.
\end{proof}

In order to describe the commutative renormalization bialgebra
$S(S(\Bcal)^+)$, we first establish some notation involving
partitions of sets.  A {\it partition} of a set $S$ is 
a set $\pi$ of nonempty, pairwise disjoint, subsets of $S$, called
the {\it blocks} of $\pi$, having union equal to $S$. We denote
by $\Pi_n$ the set of all partitions of $\{1,\dots,n\}$.
Given $a=(x_1,\dots,x_n)\in\tbn$ and a subset $B=\{i_1,\dots,i_j\}$
of $\{1,\dots,n\}$, we define $\{a|B\}\in S^j(\Bcal)$ by
$$
\{a|B\}\,=\, \{x_{i_1},\dots,x_{i_k}\}\,=\, \bigvee_{i\in B}\{x_i\},
$$
and, for $\pi=\{B_1,\dots,B_k\}\in\Pi_n$, we define
$a|\pi\in\ssb$ by
$$
a|\pi\,=\, \{a|B_1\}\cdots \{a|B_k\}\,=\, \prod_{B\in\pi}\{a|B\}.
$$
If $\Bcal$ is commutative, we regard the product
$\mub$ as a map $\sbcal^+\rightarrow B$, and in this case we
define $a/\pi\in\sbcal^+$ by
$$
a/\pi\,=\, S(\mub)(a|\pi)\,=\, \{\mub\{a|B_1\},\dots,\mub\{a|B_k\}\}.
$$
\begin{thm}\label{thm:com}
If $\Bcal$ is a commutative bialgebra then the symmetric algebra
\ssb\ is a subbialgebra of \stb.  The coproduct
of \stb, restricted to \ssb, is determined by the formula
\begin{equation}
\Delta\{a\} \,=\, \sum_{\pi\in\Pi_n}a\ix1|\pi \otimes a\ix2/\pi,
\label{comcoprod}
\end{equation}
for all $a\in\tbn$. We refer to \ssb\ as the \emph{commutative 
renormalization bialgebra}.
\end{thm}

We express the coproduct \eqref{comcoprod} analogously to
the formula \eqref{composition} for the coproduct of \ttb\
as follows:
\begin{equation}
\Delta b \,=\, \sum_{\pi\in\Pi_n} b\ix1^1\cdots b\ix1^{k}
\otimes \{\prod b\ix2^1,\dots,\prod b\ix2^k\}.
\label{partition}
\end{equation}
Here, $b=\{a\}\in S^n(\Bcal)$, where $a=(x_1,\dots,x_n)\in\tbn$ and,
for each partition $\pi\in\Pi_n$, we have $b\ix1^i=\{a\ix1|B_i\}$, and
$\prod b\ix2^i = \mub(\{a\ix2|B_i\})$, for $1\leq i\leq k$, where
$\{B_1,\dots,B_k\}$ is the set of blocks of $\pi$.\\

\noindent 
{\bf Examples:}
\begin{eqnarray*}
\Delta\{x\}\!\! &=&\! \sum \{x\ix1\} \otimes \{x\ix2\}\\
\Delta\{x,y\}\!\!&=&\! \sum \{x\ix1\}\{y\ix1\} \otimes \{x\ix2,y\ix2\}
  + \sum \{x\ix1,y\ix1\} \otimes \{x\ix2\cdot y\ix2\}\\
\Delta\{x,y,z\}\!\! &=&\! \sum \{x\ix1\}\{y\ix1\}\{z\ix1\} \otimes
   \{x\ix2,y\ix2,z\ix2\}
\\&&
+\, \sum \{x\ix1\}\{y\ix1,z\ix1\} \otimes \{x\ix2,y\ix2\cdot z\ix2\}
\\&&
+\, \sum \{y\ix1\}\{x\ix1,z\ix1\} \otimes \{y\ix2,x\ix2\cdot z\ix2\}
\\&&
+\, \sum \{z\ix1\}\{x\ix1,y\ix1\} \otimes \{z\ix2,x\ix2\cdot y\ix2\}
\\&&
+\, \sum \{x\ix1,y\ix1,z\ix1\} \otimes \{x\ix2\cdot y\ix2 \cdot z\ix2\}.\\
\Delta\{x,y,y\}\!\! &=&\! \sum \{x\ix1\}\{y\ix1\}\{y\ix1\} \otimes
   \{x\ix2,y\ix2,y\ix2\}
\\&&
+\, \sum \{x\ix1\}\{y\ix1,y\ix1\} \otimes \{x\ix2,y\ix2\cdot y\ix2\}
\\&&
+\, 2\sum \{y\ix1\}\{x\ix1,y\ix1\} \otimes \{y\ix2,x\ix2\cdot y\ix2\}
\\&&
+\, \sum \{x\ix1,y\ix1,y\ix1\} \otimes \{x\ix2\cdot y\ix2 \cdot y\ix2\}.
\end{eqnarray*}
 
\begin{proof}[Proof of Theorem \ref{thm:com}]
  
We first note that, since the sum on the right-hand side of Equation
\ref{comcoprod} is over all partitions $\pi$ of $\{1,\dots,n\}$, the
result is independent of the choice of $a\in\tbn$ representing
$\{a\}$. The fact that \ssb\ is a subbialgebra of \stb\ will follow
once we establish Equation \ref{comcoprod}, since the right-hand
side belongs to $\ssb\otimes\sbcal^+\subseteq\ssb\otimes\ssb$.

The proof depends on a basic bijection relating compositions,
permutations and partitions.  By a {\it totally ordered} partition, we
shall mean a partition $\pi$ in some $\Pi_n$ each of whose blocks is
equipped with a linear ordering, and which is linearly ordered itself.
Given a pair $(\rho,\sigma)$, where $\rho=(r_1,\dots,r_k)$ is a
composition of $n$ and $\sigma$ is a permutation of $\{1,\dots,n\}$,
we obtain a partition $\pi=\{B_1,\dots,B_k\}$ of $\{1,\dots,n\}$ by
setting
$B_i=\sigma (\rho_i)=\{\sigma(r_1+\cdots+r_{i-1}+1),\dots,
\sigma(r_1+\cdots+r_i)\}$, for $1\leq i\leq k$.  The set $\pi$ is
ordered by $B_1<\cdots<B_k$, and each $B_i$ has the linear ordering
inherited from the order $\sigma (1)<\cdots<\sigma(n)$ of
$\{1,\dots,n\}$.  The correspondence $(\rho,\sigma)\mapsto\pi$ thus
defines a bijection from the cartesian product $C_n\times\Sigma_n$
onto the set of totally ordered partitions of $\{1,\dots,n\}$.  The
inverse bijection maps a totally ordered partition $\{B_1,\dots,
B_k\}$ of $\{1,\dots,n\}$ to the pair $(\rho,\sigma)$, where $\rho
=(|B_1|,\dots,|B_k|)$, and $\sigma (i)$ is the
i{\small\it th} element of the concatenation of the linearly ordered
sets $B_1, B_2,\dots,B_k$, for $1\leq i\leq n$.

Now suppose that $a=(x_1,\dots,x_n)\in\tbn$. By definition
of $\{a\}$ and the coproduct formula \eqref{copdef}, we have
$$
\Delta\{a\}\,=\,\sum_{\rho\in C_n}\sum_{\sigma\in\Sigma_n}
a\ix1^\sigma |\rho \otimes a\ix2^\sigma /\rho,
$$
where it is understood that, for $\rho$ a composition of length $k$,
the expression $a\ix1^\sigma|\rho =
(a\ix1^\sigma|\rho_1)\cdots (a\ix1^\sigma|\rho_k)$ is the product in 
\stb. Using the commutativity of this product on the left
side of the tensor product, the commutativity of $\mub$ on the
right side, and the above bijection, we thus have
$$
\Delta\{a\} \,=\, \sum\{a\ix1|B_1\}\cdots\{a\ix1|B_k\}\otimes\{a\ix2/\pi\},
$$
where the sum is over all $\pi =\{B_1,\dots,B_k\}\in\Pi_n$; in other
words
$$
\Delta\{a\}\,=\, \sum_{\pi\in\Pi_n}a\ix1|\pi \otimes a\ix2/\pi.
$$
\end{proof} 

It is also possible to identify \ssb\ as a subspace of \ttb; that is, 
as the image
of \ttb\ under the composition 
$\alpha_{T(\mathcal B)}T(\alpha_{\mathcal B})
\colon T(T(\Bcal)) \rightarrow T(T(\Bcal))$, which maps
$u=a_1\cdots a_k\in\ttb$ to $\{\{a_1\},\dots,\{a_k\}\}\in\ssb$.
The proof of Theorem \ref{thm:com} shows that, under this
identification, the commutative renormalization
bialgebra \ssb\ is in fact a subcoalgebra of the noncommutative
renormalization bialgebra \ttb.  

\section{The Fa\`a di Bruno bialgebra}
When $\Bcal$ is the Hopf algebra of a commutative group,
there is a homomorphism from the bialgebra $S(S(\Bcal)^+)$
to the Fa\`a di Bruno algebra.
In 1855, Francesco Fa\`a di Bruno (who was beatified in 1988),
derived the general formula for the $n$th derivative of the composition
of two functions $f(g(x))$ \cite{Faa}.
In 1974, Peter Doubilet defined a bialgebra arising from
the partitions of a set \cite{Doubilet}. This bialgebra
was called the Fa\`a di Bruno bialgebra by Joni and Rota \cite{Joni},
because it is closely related to the Fa\`a di Bruno formula.
This bialgebra was further investigated by Schmitt in
\cite{Schmitt87,Schmitt94}, by Schmitt and Haiman in \cite{Haiman},
and more recently by Figueroa and Gracia-Bondia in \cite{fgcha}.

\subsection{Definition} 
\label{defFaasect}
As an algebra, the Fa\`a di Bruno bialgebra \Fcal\ is the polynomial
algebra generated by $u_n$ for $n\geq 1$. The coproduct of $\Fcal$ 
is determined by
\begin{equation}
\label{faacoprod1}
\delta u_n \,=\, \sum_{\pi\in\Pi_n} u_\pi\otimes u_{\ell (\pi)},
\end{equation}
where $u_{\pi}$ denotes the product
$\prod_{B\in\pi}u_{\scriptscriptstyle |B|}$, and
$\ell (\pi)$ the number of blocks of $\pi$,
for all partitions $\pi$.  If $\pi\in\Pi_n$ has precisely $\alpha_i$
blocks of size $i$, for all $i$, then $u_\pi=u_1^{\alpha_1}\cdots
u_n^{\alpha_n}$.  Since the number of such partitions is given by
\begin{eqnarray*}
\frac{n!}
{\alpha_1!\alpha_2!\cdots\alpha_n! (1!)^{\alpha_1}
(2!)^{\alpha_2}\cdots (n!)^{\alpha_n}}
\end{eqnarray*}
(see, e.g., \cite{Abramowitz}), it follows that the coproduct
of \Fcal\ may also be expressed as
\begin{equation}
\label{faacoprod}
\Delta u_n \,=\,
\sum_{k=1}^n
\sum_\alpha \frac{n!(u_1)^{\alpha_1} (u_2)^{\alpha_2}\cdots
(u_n)^{\alpha_n} }
{\alpha_1!\alpha_2!\cdots\alpha_n! (1!)^{\alpha_1}
(2!)^{\alpha_2}\cdots (n!)^{\alpha_n}}
\otimes u_k,
\end{equation}
where the inner sum is over the $n$-tuples of
nonnegative integers $\alpha=(\alpha_1,\alpha_2,\dots,\alpha_n)$
such that
$\alpha_1+2\alpha_2+\dots+n\alpha_n=n$
and
$\alpha_1+\alpha_2+\dots+\alpha_n=k$.
For example,
\begin{align*}
\Delta u_1 &\,=\, u_1 \otimes u_1,
\\
\Delta u_2 &\,=\, u_2 \otimes u_1 + u_1^2 \otimes u_2,
\\
\Delta u_3 &\,=\, u_3 \otimes u_1 + 3 u_1 u_2 \otimes u_2
+u_1^3 \otimes u_3,
\\
\Delta u_4 &\,=\, u_4 \otimes u_1 + 4 u_1 u_3 \otimes u_2
+ 3 u_2^2 \otimes u_2 +6 u_1^2 u_2 \otimes u_3 +u_1^4 \otimes u_4.
\end{align*}

Since \Fcal\ is a bialgebra, the set \Mcal\ of algebra maps from \Fcal\ to
the scalar field is a multiplicatively closed subset of the dual
algebra $\Fcal^*$. Using Sweedler notation $\delta u = \sum u\ix1\otimes
u\ix2$ for the coproduct of \Fcal, we have that the (convolution) product of
$f,g\in\Mcal$ is determined by $(f\star g)(u_n)=\sum f(u_n\ix1)g(u_n\ix2)$,
for all $n$.  To each element $f\in\Mcal$, we associate the
exponential series
\[
f(x) \,=\, \sum_{n=1}^\infty f(u_n)\frac{x^n}{n!}.
\]
It then follows directly from Equation \ref{faacoprod} that,
for all $f,g\in\Mcal$, the coefficient of $x^n/n!$ in
the composition $f(g(x))$ is equal to $(g\star f)(u_n)$, and thus
$f(g(x))=(g\star f)(x)$. Hence the monoid $\Mcal$ is antiisomorphic
to the monoid of all exponential formal power series
having zero constant term, under the operation of composition.


For instance, the first few coefficients of $f(g(x))$ are
\begin{align*}
(g\star f)(u_1)
 &\,=\, g_1 f_1,
\\
(g\star f)(u_2)
&\,=\, g_2 f_1 + g_1^2 f_2,
\\
(g\star f)(u_3)&
\,=\, g_3 f_1 + 3 g_1g_2 f_2  + g_1^3 f_3,
\\
(g\star f)(u_4)&
\,=\, g_4 f_1 + 4 g_1g_3 f_2  + 3g_2^2 f_2
+6g_1^2g_2 f_3+g_1^4 f_4,
\end{align*}
where we have written $f_n$ and $g_n$ for $f(u_n)$ and $g(u_n)$.

Following Connes and Moscovici \cite{ConnesM}, it is possible
to introduce a new (noncommutative) element $X$ in the algebra, such that
$[X,u_n] = u_{n+1}$, and to
generate the Fa\`a di Bruno coproduct from the relations
\begin{align*}
\Delta u_1 &\,=\, u_1\otimes u_1,\\
\Delta X &\,=\, X\otimes 1 + u_1\otimes X.
\end{align*}

\subsection{Homomorphism}
\label{Homosect}
Here, we take the bialgebra $\Bcal$ to be a commutative group Hopf algebra.
If $G$ is a commutative group, the commutative algebra $\Bcal$ is 
the vector space generated
by the elements of $G$, with product induced
by the product in $G$. The coproduct is defined by
$\deltaB x=x\otimes x$ for all elements $x\in G$.
Formula \ref{partition} for the coproduct gives
\begin{equation}
\Delta b \,=\, \sum_{\pi\in\Pi_n} b^1\cdots b^k
\otimes \{\prod b^1,\dots,\prod b^k\}.
\label{compogroup}
\end{equation}
for $b=\{x_1,\dots,x_n\}$ with $x_i\in G$.

The homomorphism $\varphi$ from $S(S(\Bcal)^+)$ to the
Fa\`a di Bruno bialgebra is given by: $\varphi(1)=1$ and
$\varphi(a)=u_n$ for any $a\in S^n(\Bcal)$ with $n>0$.  It is
clear that $\varphi$ is an algebra map; the fact that it 
respects coproducts can be established
directly by comparing Equations \ref{compogroup}
and \ref{faacoprod1}.
 
\section{The Pinter Hopf algebra}
In this section, we complete Pinter's construction, building the
noncommutative and commutative Hopf algebras that can be obtained from
the noncommutative and commutative renormalization bialgebras,
respectively.  These algebras will be called the commutative and
noncommutative Pinter Hopf algebras.

\subsection{The noncommutative case}
The renormalization bialgebra $T(T(\Bcal)^+)$ can 
be turned into a Hopf algebra by quotienting by an ideal.
The subspace $I=\{ (x)-\counitB(x) 1, x\in \Bcal\}$ of \ttb\
is a coideal because $\counit(I)=0$ and
\begin{eqnarray*}
\Delta \big((x)-\counitB(x) 1\big) &=&
\sum (x\ix1)\otimes (x\ix2) -\counitB(x) 1\otimes 1
\\ &=&
\sum (x\ix1)\otimes (x\ix2) -\sum\counitB(x\ix1) \counitB(x\ix2)1\otimes 1
\\ &=&
\sum \big((x\ix1)-\counitB(x\ix1)1\big)\otimes (x\ix2) 
\\&&
+\sum\counitB(x\ix1) 1\otimes \big((x\ix2) -\counitB(x\ix2)1\big).
\end{eqnarray*}
Therefore, the subspace $J$ of elements of the form
$u a v$, where $u,v\in \ttb$ and $a\in I$,
is an ideal and a coideal and $T(T(\Bcal)^+)/J$ is a
bialgebra \cite{Dascalescu}. 
The action of the quotient is to replace all the
$(x)$ by $\counit(x)1$. For example, we have
$\Delta(x,y) = 1 \otimes (x,y) + (x,y) \otimes 1$,
and
\begin{eqnarray*}
\Delta(x,y,z)\!\! &=&\! 1 \otimes (x,y,z)
\,+\, \sum (y\ix1,z\ix1) \otimes (x,y\ix2\cdot z\ix2)
\\&&
+\, \sum (x\ix1,y\ix1) \otimes (x\ix2\cdot y\ix2,z)
\,+\, (x,y,z) \otimes 1.
\end{eqnarray*}
More generally, if $a=(x_1,\dots,x_n)$, then
\[
\Delta a \,=\, a\otimes 1+1\otimes a
+{\sum}' a\ix1\otimes a\ix2,
\]
where $\Sigma'$ involves only elements $a\ix1$ and $a\ix2$ of degrees
strictly smaller than the degree of $a$.  Hence, 
$T(T(\Bcal)^+)/J$ is a connected Hopf algebra and
the antipode can be defined as in \cite{Schmitt87}.

\subsection{The commutative case}
The same construction can be carried out with $S(S(\Bcal)^+)/J'$, where
$J'$ is the subspace of elements of the form
$u a v$, where $u,v\in \ssb$ and 
$a\in \{ \{x\}-\counitB(x) 1, x\in \Bcal\}$.
This gives us
\begin{eqnarray*}
\Delta\{x,y\}\!\! &=&\!\! 1 \otimes \{x,y\}
  \,+\, \{x,y\} \otimes 1,\\
\Delta\{x,y,z\}\!\! &=&\!\! 1 \otimes \{x,y,z\}
+\, \sum \{y\ix1,z\ix1\} \otimes \{x,y\ix2\cdot z\ix2\}
\\&&
+\, \sum \{x\ix1,z\ix1\} \otimes \{y,x\ix2\cdot z\ix2\}
\\&&
+\, \sum \{x\ix1,y\ix1\} \otimes \{z,x\ix2\cdot y\ix2\}
\,+\, \{x,y,z\} \otimes 1.
\end{eqnarray*}
 
\subsection{The Connes-Moscovici Hopf algebra}
If we take the same quotient of the
Fa\`a di Bruno bialgebra (i.e. by letting
$u_1=1$), we obtain a Hopf algebra,
that we call the Fa\`a di Bruno Hopf algebra.
In the course of a proof of index theorems
in noncommutative geometry, Connes and Moscovici
defined a noncommutative Hopf algebra \cite{ConnesM}.
They noticed that the commutative part of this
Hopf algebra is related to the algebra of
diffeomorphisms as follows:
If $\phi(x)=x+\sum_{n=2}^\infty u_n x^n/n!$,
they define $\delta_n$ for $n>0$ by
\[
\log \phi'(x) \,=\, \sum_{n=1}^\infty \delta_n \frac{x^n}{n!}.
\]
To calculate $\delta_n$ as a function of 
$u_k$, we use the fact that
$\phi'(x)=1+\sum_{n=1}^\infty u_{n+1} x^n/n!$
and
$\log(1+z)=\sum_{k=1}^\infty (-1)^{k-1} (k-1)! z^k/k!$.
Since the Fa\`a di Bruno formula describes the composition
of series, we can use it to write immediately
\[
\delta_n \,=\,
\sum_{k=1}^n (-1)^{k-1} (k-1)!
\sum_\alpha \frac{n!(u_2)^{\alpha_1} (u_3)^{\alpha_2}\cdots
(u_{n+1})^{\alpha_n} }
{\alpha_1!\alpha_2!\cdots\alpha_n! (1!)^{\alpha_1}
(2!)^{\alpha_2}\cdots (n!)^{\alpha_n}},
\]
where the sum is over the $n$-tuples of
nonnegative integers $\alpha=(\alpha_1,\alpha_2,\dots,\alpha_n)$
such that
$\alpha_1+2\alpha_2+\dots+n\alpha_n=n$
and
$\alpha_1+\alpha_2+\dots+\alpha_n=k$.
 
For example, $\delta_1 = u_2$, $\delta_2 = u_3-u_2^2$,
$\delta_3 = u_4-3u_3u_2+2u_2^3$.
Note that, except for the shift, the relation between
$u_n$ and $\delta_n$ is the same as the
relation between the moments of a distribution and
its cumulants, or between unconnected
Green functions and connected Green functions. 
The inverse relation is obtained from
\[
\phi(x) \,=\, \int_0^x \dd t \exp\big(\sum_{n=1}^\infty 
\delta_n \frac{t^n}{n!}\big);
\]
thus
\[
u_{n+1} \,=\,
\sum_\alpha \frac{n!(\delta_1)^{\alpha_1}\cdots (\delta_{n})^{\alpha_n} }
{\alpha_1!\cdots\alpha_n! (1!)^{\alpha_1}\cdots (n!)^{\alpha_n}}.
\]
There is also a morphism between the noncommutative Pinter Hopf algebra
and the noncommutative algebra of diffeomorphisms, which was defined
in \cite{BFII}.   The relation between the Fa\`a di Bruno
Hopf algebra and the Connes-Moscovici algebra was also studied
in \cite{fgcha}.

\section{Relation with renormalization}
The renormalization of time-ordered products in configuration space
was first considered by Bogoliubov, Shirkov and Parasiuk in
\cite{BS55,BS56,BP57} and presented in detail in the textbook
\cite{Bogoliubov}. It was elaborated more precisely in \cite{Epstein}
and received its Hopf algebraic formulation in \cite{PinterHopf}.
This approach was found particularly convenient for defining quantum
field theories in curved spacetime
\cite{Brunetti,Brunetti2,Hollands,Hollands4}.  Here we show that the
renormalization defined by Bogoliubov and Shirkov (see
\cite{Bogoliubov}, section 26) or by Pinter \cite{PinterHopf} can be
obtained from our construction.  We first define the bialgebra
of fields $\Bcal$.

\subsection{The bialgebra of fields}
\label{bialgebrafieldssect}
We consider a finite set $\calD$ of distinct points in $\mathbb{R}^d$.
The bialgebra $\Bcal$ is generated as a vector space
over $\mathbb{C}$
by the symbols $\phi^n(x)$, where $n$ is a nonnegative
integer and $x\in\calD$.
The basis elements $\phi^n(x)$ are called Wick monomials
in the physical literature.
The algebra product of $\Bcal$ is defined by
$\phi^n(x) \cdot \phi^m(y) = \delta_{x,y} \phi^{n+m}(x)$
and its unit is $1_{\mathcal B} =\sum_{x\in\mathcal D} \phi^0(x)$.

The coproduct of $\Bcal$ is the binomial coproduct \cite{Majid}
\[
\deltaB \phi^n(x) \,=\, \sum_{k=0}^n \binom{n}{k}
\phi^k(x) \otimes \phi^{n-k}(x),
\]
and the counit is $\counitB(\phi^n(x))=\delta_{n,0}$.
We define the degree of $\phi^n(x)$ to be $n$.
The algebra $\Bcal$ is thus a graded commutative and cocommutative
bialgebra. It is infinite dimensional but finite
dimensional in each degree.

The symmetric algebra $S(\Bcal)$ has coproduct $\delta$ 
induced by the coproduct of $\Bcal$. 
More explicitly, the coproduct of 
$a=\{\phi^{n_1}(x_1),\dots,\phi^{n_m}(x_m)\}$ is
\begin{eqnarray*}
\delta a &=&
\sum_{i_1=0}^{n_1}\cdots\sum_{i_m=0}^{n_m} 
\binom{n_1}{i_1}\cdots\binom{n_m}{i_m}
\\&&
\{\phi^{i_1}(x_1),\dots,\phi^{i_m}(x_m)\}
\otimes
\{\phi^{n_1-i_1}(x_1),\dots,\phi^{n_m-i_m}(x_m)\}.
\end{eqnarray*}
The counit is defined by $\counitT(1)=1$
and $\counitT(\phi^n(x))=\delta_{n,0}$, and extended to
$S(\Bcal)$ by linearity and multiplicativity.
This coproduct and counit turn $S(\Bcal)$ 
into a commutative and cocommutative biagebra.

{\bf {Remarks}}:
(i) If we restrict the definition to a single point $x$,
the commutative and cocommutative algebra $\Bcal$ can be identified 
with the algebra of Wick monomials at $x\in\calD$
\cite{Brunetti,Brunetti2}.
(ii) The product in $S(\Bcal)$ is the usual normal product of
quantum field theory \cite{ReedSimonII}. 
In other words,
$\{\phi^{n_1}(x_1),\dots,\phi^{n_m}(x_m)\}$ would be
written
${:}\phi^{n_1}(x_1)\cdots\phi^{n_m}(x_m){:}$ in a quantum
field theory textbook. The product in $S(\Bcal)$ can
also be considered as a product of classical fields \cite{Dutsch04}.
(iii) In quantum field theory, the counit of $S(\Bcal)$
is called the vacuum expectation value
\cite{BrouderOecklI}:
$\counitT(a)=\langle 0|a|0\rangle$ for 
$a\in S(\Bcal)$.
(iv) Considering a
finite number of points $\calD$ instead of all
the points of $\mathbb{R}^d$ is consistent 
with the framework of perturbative renormalization
and with the fact that the renormalization Hopf algebra
encapsulates the combinatorics of renormalization but not
its analytical aspects.
Taking all the points of $\mathbb{R}^d$ would
make $\Bcal$ a non-locally-compact Hopf algebra, i.e.
an object very difficult to handle.

\subsection{Time-ordered product}
To define the time-ordered product, we start from
a linear map $t: S(\Bcal)\rightarrow \mathbb{C}$
such that $t(1)=1$.
The time-ordered product $T$ is a linear map
$S(\Bcal) \rightarrow S(\Bcal)$ defined by
\begin{equation}
T(a) \,=\, \sum t(a\ix1) a\ix2, 
\label{T(a)}
\end{equation}
Note that $t(a)$ can be recovered from $T(a)$
by the relation $t(a)=\counitT\big(T(a)\big)$.

{\bf {Remarks}}:
(i)
In quantum field theory, the map $t$ is defined
in terms of Feynman diagrams \cite{BrouderGroup24},
but the combinatorics of renormalization does not depend
on the precise structure of $t$.
(ii)
Equation \ref{T(a)} was essentially given
in the paper by Epstein and Glaser \cite{Epstein}.
In the physics literature, it is written
\cite{Brunetti,Epstein}
\begin{eqnarray*}
T\big(\phi^{n_1}(x_1)\cdots\phi^{n_m}(x_m)\big)\!\!
&=&\! \sum_{i_1=0}^{n_1}\cdots\sum_{i_m=0}^{n_m} 
\binom{n_1}{i_1}\cdots\binom{n_m}{i_m}
\\&&\hspace*{-30mm}
\langle 0|T\big(\phi^{i_1}(x_1)\cdots\phi^{i_m}(x_m)\big)
|0\rangle
{:}\phi^{n_1-i_1}(x_1)\cdots\phi^{n_m-i_m}(x_m){:}.
\end{eqnarray*}

\subsection{Relation between time-ordered products}
According to Bogoliubov and Shirkov \cite{Bogoliubov}, renormalization
can be seen as a particular kind of transformation
from a time-ordered product $T$ defined by a map $t$
to a time-ordered product $\tilde T$ defined by a map $\tilde t$.
If $a=\{\phi^{n_1}(x_1),\dots,\phi^{n_m}(x_m)\}$,
in standard quantum field theory 
$t(a)$ is defined in terms of regularized Feynman propagators
and $\tilde t(a)$ supplies the counterterms that remove the
singularities of $t(a)$; in the Epstein-Glaser approach
the Feynman propagators are not regularized and
$t(a)$ is well-defined only
when all spacetime points $x_i$ are different, then
$\tilde t$ is the extension of $t$ to the case
of coinciding spacetime points.
It is also common in physics to consider renormalization
where $t$ and $\tilde t$ are both well-defined.
This finite renormalization determines the effect of a change of
the parameters (mass, coupling constant) describing the
physical system.

The relation between time-ordered products $T$ and $\tilde T$ was
discussed in 
\cite{PinterHopf,Bogoliubov,Epstein,Brunetti2,Hollands,Hollands4,Pinter}.
We shall follow the presentation given by Pinter \cite{PinterHopf}.
She considers linear maps $O : S(\Bcal)\rightarrow \Bcal$.
Apart from linearity, the only specific property of the maps $O$
is the fact that they are diagonally supported. 
This expresses the local nature of renormalization and means that
$O\big(\{\phi^{n_1}(x_1),\dots,\phi^{n_m}(x_m)\}\big)$
is zero if the relation $x_1=x_2=\cdots=x_m$ is not satisfied.
The other properties of $O$ will be consequences of the
fact that $T$ and $\tilde T$ are time-ordered products.
In \cite{PinterHopf}, Pinter uses the notation
$\Delta$ for our $O$, but we changed to
$O$ (as in \cite{Hollands4}) to avoid confusion
with the coproduct. The elements $O(a)$ are called
\emph{quasilocal operators} and denoted by
$\Lambda$ or $\Delta$ by
Bogoliubov and Shirkov \cite{Bogoliubov}.
The purpose of this section is a description of $O(a)$
in terms of the coproduct of $S(\Bcal)$ and to
show how the definition of $O$ must be modified
to make it consistent with our algebraic approach.

The map $O$ is used to define $\tilde T$ from $T$.
Equation 13 of Pinter's paper \cite{PinterHopf} can be written
(see also \cite{Bogoliubov})
\begin{equation}
\tilde{T}(a) \,=\, \sum_{\pi\in\Pi_m} 
   T\big(O(b^1)\pvee\cdots\pvee O(b^k)\big),
\label{tildeT}
\end{equation}
where we use the notation of
Equation \ref{partition}:
if $a=\{\phi^{n_1}(x_1),\dots,\phi^{n_m}(x_m)\}$ is an element of
$S(\Bcal)$, we let $b=(\phi^{n_1}(x_1),\dots,\phi^{n_m}(x_m))\in\tb$,
and for $\pi$ a partition with blocks $B_1,\dots B_k$, we have
$b^i = \{b|B_i\}$, for $1\leq i\leq k$.
For notational convenience, we identify $S^1(\Bcal)$ with
$\Bcal$ in the rest of the section.

We derive now some additional properties of $O$.
In standard quantum field theory, a single vertex
is not renormalized \cite{Bogoliubov}. 
Thus $\tilde{T}(a)=T(a)=a$ if $a\in \Bcal$.
This enables us to show that $O(a)=a$ if
$a\in \Bcal$: If $a\in \Bcal$, then the sum in
Equation \ref{tildeT} has only one term, corresponding
to the partition $\{\{1\}\}$ of the set $\{1\}$, and thus
$\tilde{T}(a)=T\big(O(a)\big)$.
But $O(a)\in \Bcal$ (by definition of $O$), and thus $T\big(O(a)\big)=O(a)$.
The fact that $\tilde{T}(a)=a$ implies that $O(a)=a$.

If $a=\{\phi^{n_1}(x_1),\dots,\phi^{n_m}(x_m)\}$
with $m>1$, we use $O(\{\phi^{n_i}(x_i)\})=\{\phi^{n_i}(x_i)\}$
to rewrite Equation \ref{tildeT}
\begin{equation*}
\tilde{T}(a) \,=\, T(a)+ T(O(a))+{\sum_{\pi\in\Pi_m'}}
 T\big(O(b^1)\pvee\cdots\pvee O(b^{k})\big),
\end{equation*}
where $\Pi_m'$ indicates set of all partitions of
$\{1,\dots, m\}$, except for
$\pi = \{\{1\},\dots,\{m\}\}$ and $\pi = \{\{1,\dots,m\}\}$.
But $O(a)\in \Bcal$ and $T$ acts as the identity on $\Bcal$,
thus
\begin{equation}
\tilde{T}(a) \,=\, T(a)+ O(a)+{\sum_{\pi\in\Pi_m'}} 
 T\big(O(b^1)\pvee\cdots\pvee O(b^{k})\big).
\label{tildeT2}
\end{equation}

From this transformation formula and the support property of
$O$ we deduce 

\begin{pro}
\label{propO}
If $a=\{\phi^{n_1}(x_1),\dots,\phi^{n_m}(x_m)\}$, then
$O(a)=\sum c(a\ix1)a\ix2$, where
$c(a)=\counitT\big(O(a)\big)$. Moreover,
if $m=1$, then $c(a)=\counitT(a)$,
if $m>1$, then $c(a)$ is supported on
$x_1=\cdots=x_m$ and can be obtained recursively from
$\tilde{t}$ and $t$ by
\begin{equation}
c(a) = \tilde{t}(a)-t(a) -{\sum_{\pi\in\Pi_m'}}\sum
c(b\ix1^1)\cdots c(b\ix1^k) t(b\ix2^1\pvee\cdots\pvee b\ix2^k).
\label{defc}
\end{equation}
\end{pro}
\begin{proof}
The proof is by induction.
Take $a=\{\phi^{n_1}(x_1),\dots,\phi^{n_m}(x_m)\}$.
If $m=1$, we have $O(a)=a$, so that
$O(a)=\sum \counitT(a\ix1) a\ix2$ and $c(a)=\counitT(a)$.
For $m=2$, the set $\Pi_m'$ is empty, and
thus Equation \ref{tildeT2} yields
$\tilde{T}(a)=T(a)+O(a)$. We know that
$T(a)=\sum t(a\ix1) a\ix2$ and
$\tilde{T}(a)=\sum \tilde{t}(a\ix1) a\ix2$,
thus $O(a)=\sum c(a\ix1) a\ix2$
with $c=\tilde{t}-t$.
Now assume that the proposition is true up to
$m-1$ and take $a=\{\phi^{n_1}(x_1),\dots,\phi^{n_m}(x_m)\}$.
In Equation \ref{tildeT2}, 
all $O(b^i)$ can be written $\sum c(b\ix1^i)b\ix2$
because the degree of $b^i$ is smaller than $m$.
Hence, we can write
\begin{align*}
\tilde{T}(a) &\,=\, 
T(a)\,+\, O(a)\,+\,{\sum_{\pi\in\Pi_m'}}\sum c(b\ix1^1)\cdots c(b\ix1^k)
T\big(b\ix2^1\pvee\cdots\pvee b\ix2^k\big).
\end{align*}
Equation \ref{T(a)} now yields
\begin{eqnarray*}
\sum \tilde{t}(a\ix1) a\ix2\!\! &=&\!\! \sum t(a\ix1) a\ix2\,+\, O(a)
\\&&\hspace*{-5mm}+\,
{\sum_{\pi\in\Pi_m'}}\sum c(b\ix1^1)\cdots c(b\ix1^k)
t\big(b\ix2^1\pvee\cdots\pvee b\ix2^k\big)
b\ix3^1\pvee\cdots\pvee b\ix3^k.
\end{eqnarray*}
Since $a=b^1\pvee\cdots\pvee b^k$, the factor
$b\ix3^1\pvee\cdots\pvee b\ix3^k$ can be written as $a\ix3$, and
Equation \ref{defc} follows from  the coassociativity
of the coproduct.

The fact that $O(a)$ is supported on
$x_1=\cdots=x_m$ implies that $c(a)=\counitT\big(O(a)\big)$
is supported on $x_1=\cdots=x_m$. 
Thus, 
$c(a)=f(x_1) \delta_{x_2,x_1}\cdots \delta_{x_m,x_1}$,
where $f$ is some function of $x_1$.
In flat spacetime and in the absence of an external field,
the system is translation invariant and $f(x_1)$
is a constant. 
\end{proof}

Now comes a crucial step which is not apparent in the usual
renormalization.
In quantum field theory, when $x_1=\dots=x_m$,
$\{\phi^{n_1}(x_1),\dots,\phi^{n_m}(x_m)\}$
is identified with
\begin{eqnarray*}
\prod_{p=1}^m \phi^{n_p}(x_1)
&=& \phi^{n_1+\cdots+n_m}(x_1),
\end{eqnarray*}
where the product $\prod$ is in $\Bcal$.
After of this identification, the expression
$O(a)=\sum c(a\ix1) a\ix2$ is replaced by
\begin{eqnarray}
\Lambda(a)=\sum c(a\ix1) \prod a\ix2,
\label{defLambda}
\end{eqnarray}
where the product means that, if 
$a=y_1\pvee y_2\pvee\cdots\pvee y_p$ with $y_i\in\Bcal$,
then $\prod a= y_1\cdot y_2 \cdots y_p$ where the
product $\cdot$ is in $\Bcal$.
From our algebraic point of view, $\Lambda(a)$ is different from $O(a)$.
The map $\Lambda$ is more satisfactory because
$\prod a\ix2$ belongs to $\Bcal$ and it is clear
that $\Lambda$ maps $S(\Bcal)$ to $\Bcal$.
This was not the case with the
expression $O(a)=\sum c(a\ix1) a\ix2$
because $a\ix2$ belongs to $S(\Bcal)$.

Therefore, in the following, we shall use
$\Lambda$ instead of $O$ to define
the renormalized time-ordered product $\barT$
as the linear operator $S(\Bcal)\rightarrow S(\Bcal)$
\begin{equation}
\barT(a) \,=\, \sum_\pi 
  T\big(\Lambda(b^1)\pvee\cdots\pvee \Lambda(b^{k})\big).
\label{barT}
\end{equation}
If we expand the terms $\Lambda(b^i)$ with Equation
\ref{defLambda} we find
\begin{equation}
\barT(a) = 
\sum_\pi \sum c(b\ix1^1)\cdots c(b\ix1^k) 
  T(\prod b\ix2^1\pvee\cdots\pvee\prod b\ix2^k).
\label{barTform}
\end{equation}
Although $\barT(a)$ would be undistinguishable from
$\tilde{T}(a)$ in quantum field theory, these two quantities
are different in our algebraic approach.
The main difference is the fact that there is no
map $\bart$ such that $\barT(a) = \sum \bart(a\ix1) a\ix2$.

Equation \ref{barTform} can be translated into
the usual renormalization prescription by saying that
each $b^i$ is a generalized vertex in the sense of
Bogoliubov and Shirkov \cite{Bogoliubov}, the operation
$\prod b\ix2^i$ shrinks the points of $b\ix2^i$ into
a single point leaving the external lines unchanged
and the number $c(b\ix1^i)$ describes the counterterm
associated to the generalized vertex.

\subsection{Relation to the renormalization coproduct}
It remains to relate the last result to the
renormalization coproduct.
If we compare expression \ref{barTform} to the commutative
renormalization coproduct \eqref{partition}, we see that
\begin{equation}
\barT(a) \,=\, \sum C(a\ii1) T(a\ii2),
\label{renormT}
\end{equation}
where $C(a)=c(a)$ for $a\in S(\Bcal)^+$ and $C(uv)=C(u)C(v)$ for
$u,v\in S(S(\Bcal)^+)$, and where we have used the alternate Sweedler
notation $\Delta (a) = \sum a\ii1\otimes a\ii2$ for the commutative
renormalization coproduct. 
The quantum field relation
$\tilde T(a)=\sum \tilde t(a\ix1)a\ix2$ becomes
\begin{equation*}
\barT(a) \,=\, \sum C(a\ii1) t(a\ii2\ix1)a\ii2\ix2.
\end{equation*}

Renormalization can be seen from (at least) two points of view:
(i) as a way to transform a time-ordered product $T$
into a renormalized time-ordered product $\barT$,
(ii) as a way to transform a bare Lagrangian
(i.e. an element $a$ of $\Bcal$) into a renormalized
Lagrangian.
We establish now the connection between these two points of view.
If $a\in\Bcal$ we define $a^n\in S^n(\Bcal)$ by
$a^n=a\pvee \cdots\pvee a$ ($n$ times).
With this notation we can define, in the sense of formal
power series in a complex variable $\lambda$, the series
$\ee^{\lambda a}=\sum \lambda^n a^n /n!$.
We have
\begin{pro}
\label{prop5}
If $a$ is an element of $\Bcal$
and $\lambda$ a complex variable, then
\begin{equation}
\barT\big(\ee^{\lambda a}\big)= T\big(\ee^{\lambda \bara(\lambda)}\big),
\label{barTS}
\end{equation}
where
\begin{equation}
\bara(\lambda) = \sum_{n=1}^\infty \frac{\lambda^{n-1}}{n!}
  \Lambda(a^n)
= \Lambda\Big(\frac{\ee^{\lambda a}-1}{\lambda}\Big).
\label{defRL}
\end{equation}
\end{pro}
In this proposition, Equations \ref{barTS} and $\ref{defRL}$
are understood in the sense of formal power series in $\lambda$.
The first term of $\bara(\lambda)$ is $\Lambda(a)=a$,
which is called the bare Lagrangian in quantum field theory.
The next terms are called the counterterms of the Lagrangian.
The proof of the proposition is straighforward:
\begin{proof}
We expand $\barT\big(\ee^{\lambda a}\big)$
as
\begin{equation*}
\barT\big(\ee^{\lambda a}\big)= 1
+ \sum_{n=1}^\infty \frac{\lambda^n}{n!} \barT(a^n).
\end{equation*}
To calculate $\barT(a^n)$ we use Equation \ref{barTform}, noting
that in this case the elements $b^1,\dots,b^k\in\sbcal$ depend only
on the sizes of the blocks of $\pi$. 
The number of partitions
of $n$ different objects with $\alpha_i$ blocks of
size $i$ was given in Section \ref{Homosect}. This gives us
\begin{equation*}
\barT(a^n) = \sum_\alpha
\frac{n!
T\big(\Lambda(a^1)^{\alpha_1}\pvee\cdots\pvee
\Lambda(a^n)^{\alpha_n}\big)}
{\alpha_1!\alpha_2!\cdots\alpha_n! (1!)^{\alpha_1}
(2!)^{\alpha_2}\cdots (n!)^{\alpha_n}},
\end{equation*}
where the $n$-tuples $\alpha$ are described in section \ref{defFaasect}.
Consider now
$g(\lambda)=\lambda \bara(\lambda)$. This is a formal exponential power
series with coefficients $g_n=\Lambda(a^n)$.
Thus, 
$ T\big(\ee^{\lambda \bara(\lambda)}\big)=
1+ T\big(\ee^{g(\lambda)}-1\big)= 1+T\big(f(g(\lambda))\big)$, where
$f(\lambda)=\ee^\lambda -1$ is an exponential power series
with coefficients $f_n=1$. If we use the Fa\`a di Bruno
formula for the composition of series, we obtain the term
of degree $\lambda^n$ in the exponential series $f(g(\lambda))$ as
\begin{equation*}
\sum_\alpha
\frac{n!
\Lambda(a^1)^{\alpha_1}\pvee\cdots\pvee
\Lambda(a^n)^{\alpha_n}}
{\alpha_1!\alpha_2!\cdots\alpha_n! (1!)^{\alpha_1}
(2!)^{\alpha_2}\cdots (n!)^{\alpha_n}}.
\end{equation*}
Therefore, the linearity of $T$ implies that
\begin{equation*}
\barT\big(\ee^{\lambda a}\big)= 1
+ \sum_{n=1}^\infty \frac{\lambda^n}{n!} \barT(a^n)
= 1 + T\big(f(g(\lambda))\big) = 
T\big(\ee^{\lambda \bara(\lambda)}\big).
\end{equation*}
\end{proof}

We have thus shown that the renormalization coproduct
gives the same result as the 
Bogoliubov-Shirkov-Epstein-Glaser renormalization.
On the other hand, it was proved in \cite{Epstein}
and \cite{PinterPhD} that the latter renormalization
coincides with the standard BPHZ renormalization.
Thus, the renormalization coproduct recovers the standard
renormalization of scalar fields.

The coassociativity of the renormalization coproduct
means that the transitions from one renormalized time-ordered
product to another can be composed associatively.
In other words, if $\barT(a) = \sum C(a\ii1) T(a\ii2)$
and $\barT'(a) = \sum C'(a\ii1) \barT(a\ii2)$, then 
$\barT'(a) = \sum C''(a\ii1) T(a\ii2)$ 
with $C''(a)= \sum C'(a\ii1) C(a\ii2)$.

In this section we have assumed that
$T\big(\phi^n(x)\big)=\phi^n(x)$.
This is consistent with the use of the Pinter algebra.
It was pointed out by Hollands and
Wald \cite{Hollands,Hollands4} that
renormalization at a point is necessary
in curved spacetime and is given by
\[
\Lambda(\phi^n(x)) \,=\, \sum_{k=0}^n \binom{n}{k}
c\big(\phi^{k}(x)\big) \phi^{n-k}(x).
\]
This is exactly Equation \ref{defLambda}.
Thus, the renormalization of quantum field
theory in curved spacetimes requires 
the commutative renormalization
bialgebra instead of the Pinter Hopf algebra.

Many equations of this section are valid for any commutative
bialgebra $\Bcal$: the definition \eqref{T(a)} of the
time-ordered product is the right coregular action of
$S(\Bcal)^*$ on $S(\Bcal)$; for $c\in S(\Bcal)^*$,
Equation \ref{defLambda} defines
a map $\Lambda : S(\Bcal)\rightarrow \Bcal$, and
the renormalized time-ordered product
\eqref{barT} is then well defined and satisfies Proposition
\ref{prop5}.

{\bf{Remark}}. 
If we take the realistic example of a quantum field theory
with the interaction $a=\int\phi(x)^4\dd x$
in four spacetime dimensions (in our framework, the
integral should be replaced by a finite sum),
the quasilocal operators take the explicit form
\cite{Bogoliubov}:
\begin{eqnarray*}
\Lambda(a^n) &=& C^n_1 \int \phi^2(x) \dd x
+C^n_2 \int \phi^4(x)\dd x
+ C^n_3 \int \phi(x)\Box\phi(x) \dd x,
\end{eqnarray*}
where $C^n_i$ are constants.
In this equation, the first two quasilocal operators
renormalize logarithmic divergences and are of the type
treated in this paper. The third quasilocal operator
involves derivatives of the fields. It is used to
remove quadratic divergences and is absent in our approach.
In other words, the present renormalization algebra can only
deal with logarithmic divergences.
To take care of higher divergences we should need to study
the interplay between the renormalization algebra
and derivations of this algebra.
However, even at the quantum field level, the interplay between
time-ordered products and derivatives
is a delicate matter \cite{Dutsch02,Dutsch03,Dutsch04}.
Its Hopf algebraic interpretation is still an
open problem.
Within the Connes-Kreimer approach, this problem was
solved in terms of an ``external structure'' \cite{CKI}.

\section{Conclusion}
We have constructed the renormalization bialgebra 
corresponding to any bialgebra $\Bcal$. 
In standard quantum field
theory, the commutative Pinter Hopf algebra is
generally used, but the noncommutative one may
be relevant to the renormalization of some
noncommutative quantum field theories
\cite{Ritter}.

At the mathematical level, the renormalization
Hopf algebra found unexpected applications in 
number theory \cite{Fauser06}. Moreover,
an intriguing connection was observed between 
the renormalization bialgebra \ttb\ and a construction 
involving operads \cite{MoerdijkvanderLaan,vanderLaan}.
Such a connection is 
another manifestation of the deep mathematical meaning 
of renormalization.
This operad is a realization of
Kreimer's suggestion that operads should play
a role in renormalization theory \cite{KreimerPC}.
Kreimer himself defined an operad of renormalization
based on Feynman diagrams \cite{KreimerPR}.

\section{Acknowledgements}
We are very grateful to Gudrun Pinter for the detailed
information on her results that she kindly gave us.
We wish to thank Alessandra Frabetti, Muriel Livernet,
Klaus Fredenhagen, Romeo Brunetti, Robert Oeckl, Jos\'e M. Gracia-Bondia,
Michael D\"utsch, Ieke Moerdijk, Pepijn van der Laan, 
Bertfried Fauser, Fr\'ed\'eric Patras and Christian Voigt for 
helpful comments and discussions.  
A special thanks to Jean-Louis Loday for his
constant help and encouragement.  Finally,
we thank the referee for many valuable comments and
suggestions.

\providecommand{\bysame}{\leavevmode\hbox to3em{\hrulefill}\thinspace}
\providecommand{\MR}{\relax\ifhmode\unskip\space\fi MR }
\providecommand{\MRhref}[2]{%
  \href{http://www.ams.org/mathscinet-getitem?mr=#1}{#2}
}
\providecommand{\href}[2]{#2}


\begin{thebibliography}{10}

\bibitem{Abramowitz}
M.~Abramowitz and I.A. Stegun, \emph{Handbook of mathematical 
functions, with formulas, graphs, and mathematical tables}, 
fifth ed., Dover, New York, 1964.

\bibitem{Bogoliubov}
N.N. Bogoliubov and D.V. Shirkov, \emph{Introduction to the theory of 
quantized fields}, Interscience Pub. Inc., New York, 1959.

\bibitem{BS55}
N.N. Bogoliubov and D.W. Shirkov, \emph{(in {R}ussian)}, Uzpekhi fiz. Nauk
\textbf{57} (1955), 3--91.

\bibitem{BP57}
N.N. Bogoljubow and O.S. Parasiuk, \emph{{\"U}ber die {M}ultiplikation der
{K}ausalfunktionen in der {Q}uantentheorie der {F}elder}, Acta Math.
\textbf{97} (1957), 227--266.

\bibitem{BS56}
N.N. Bogoljubow and D.W. Schirkow, \emph{Probleme der {Q}uantenfeldtheorie.
{II. B}eseitigung der {D}ivergenzen aus der {S}treumatrix}, Fort. d. Phys.
\textbf{4} (1956), 438--517.

\bibitem{BrouderGroup24}
Ch. Brouder, \emph{Quantum groups and interacting quantum fields}, 
Group 24: Physical and Mathematical Aspects of Symmetries
(Bristol) (J.-P. Gazeau, R.~Kerner, J.-P. Antoine, 
S.~M{\'e}tens, and J.-Y. Thibon, eds.), Institute of Physics Publishing, 
2003, pp.~787--790.

\bibitem{BrouderQG}
Ch. Brouder, B.~Fauser, A.~Frabetti, and R.~Oeckl, \emph{Quantum field theory
and {H}opf algebra cohomology}, J. Phys. A: Math. Gen. \textbf{34} (2004),
5895--5927.

\bibitem{BFII}
Ch. Brouder, A.~Frabetti, and Ch. Krattenthaler, \emph{Noncommutative {H}opf
algebra of formal diffeomorphisms}, Adv. Math. \textbf{200} (2006), 
479-524.

\bibitem{BrouderOecklI}
Ch. Brouder and R.~Oeckl, \emph{Quantum groups and quantum field theory: The
free scalar field}, Mathematical Physics Research on the Leading Edge, 
(Hauppauge NY)
(C.V. Benton, ed.), Nova Science, 2004, pp.~63--90.

\bibitem{Brunetti}
R.~Brunetti and K.~Fredenhagen, \emph{Interacting quantum fields in curved
space: {R}enormalization of {$\phi^4$}}, Operator Algebras and Quantum Field
Theory (Cambridge) (S.~Doplicher, R.~Longo, J.E. Roberts, and L.~Zsido,
eds.), International Press of Boston, 1997, pp.~546--563.

\bibitem{Brunetti2}
\bysame, \emph{Microlocal analysis and interacting quantum field theories:
renormalization on physical backgrounds}, Commun. Math. Phys. \textbf{208}
(2000), 623--661.

\bibitem{CKI}
A.~Connes and D.~Kreimer, \emph{Renormalization in quantum field theory 
and the {R}iemann-{H}ilbert problem {I}: the {H}opf algebra structure of 
graphs and the main theorem}, Commun. Math. Phys. \textbf{210} (2000), 
249--273.

\bibitem{ConnesM}
A.~Connes and H.~Moscovici, \emph{Hopf algebras, cyclic cohomology and the
transverse index theorem}, Commun. Math. Phys. \textbf{198} (1998), 199--246.

\bibitem{Dascalescu}
S.~D{\u a}sc{\u a}lescu, C.~N{\u a}st{\u a}sescu, and {\c S}.~Raianu,
\emph{Hopf algebras}, Marcel Dekker Inc., New York, 2001.

\bibitem{Faa}
F.~Fa{\`a} di~Bruno, \emph{Sullo svilupo delle funzioni}, Ann. Sci. Mat. Fis.
\textbf{6} (1855), 479--480.

\bibitem{Doubilet}
P.~Doubilet, \emph{A {H}opf algebra arising from the lattice of partitions 
of a set}, J. Alg. \textbf{28} (1974), 127--132.

\bibitem{Dutsch02}
M.~D{\"u}tsch and F.-M. Boas, \emph{The master {W}ard identity}, Rev. Math.
Phys. \textbf{14} (2002), 977--1049.

\bibitem{Dutsch03}
M.~D{\"u}tsch and K.~Fredenhagen, 
\emph{The master {W}ard identity and generalized {S}chwinger-{D}yson
equation in classical field theory}, Commun. Math. Phys. \textbf{243} (2003),
275--314.

\bibitem{Dutsch04}
\bysame, \emph{Causal perturbation theory in terms of retarded
products, and a proof of the {A}ction {W}ard {I}dentity}, Rev. Math.
Phys. \textbf{16} (2004), 1291--1348.

\bibitem{Epstein}
H.~Epstein and V.~Glaser, \emph{The role of locality in perturbation theory},
Ann. Inst. Henri Poincar\'e \textbf{19} (1973), 211--295.

\bibitem{Fauser06}
B.~Fauser, \emph{The Dirichlet Hopf algebra of arithmetics},
J. Knot Theory Ramifications \textbf{16} (2007), 1--60.

\bibitem{fgcha}
H.~Figueroa and J.M. Gracia-Bondia, \emph{Combinatorial {H}opf algebras in
quantum field theory {I}}, Rev. Math.
Phys. \textbf{17} (2005), 881--976.

\bibitem{GraciaLazzarini}
J.M. Gracia-Bondia and S.~Lazzarini, 
\emph{Connes-{K}reimer-{E}pstein-{G}laser renormalization},  
(2000), arXiv:hep-th/0006106.

\bibitem{GVF}
J.M. Gracia-Bondia, J.C. Varilly, and H.~Figueroa, \emph{Elements of
noncommutative geometry}, Birkh{\"{a}}user, Boston, 2001.

\bibitem{Haiman}
M.~Haiman and W.~Schmitt, 
\emph{Incidence algebra antipodes and {L}agrange
inversion in one and several variables}, J. Combin. Theor. A \textbf{50}
(1989), 172--185.

\bibitem{HivertPhD}
F.~Hivert, \emph{Combinatoire des fonctions quasi-sym{\'e}triques}, Ph.D.
thesis, Marne-la-Vall{\'e}e University, 1999.

\bibitem{Hollands}
S.~Hollands and R.M. Wald, \emph{Local {W}ick polynomials and time ordered
products of quantum fields in curved spacetime}, Commun. Math. Phys.
  \textbf{223} (2001), 289--326.

\bibitem{Hollands4}
\bysame, \emph{On the renormalisation group in curved spacetime}, 
Commun. Math. Phys. \textbf{237} (2003), 123--160.

\bibitem{Joni}
S.A. Joni and G.-C. Rota, \emph{Coalgebras and bialgebras in combinatorics},
Stud. Appl. Math. \textbf{61} (1979), 93--139.

\bibitem{Kassel}
Ch. Kassel, \emph{Quantum groups}, Springer Verlag, New York, 1995.

\bibitem{KreimerPC}
D.~Kreimer, 1999, Personal communication.

\bibitem{KreimerPR}
\bysame, \emph{Combinatorics of (perturbative) quantum field theory}, Phys.
Repts. \textbf{363} (2002), 387--424.

\bibitem{Majid}
S.~Majid, \emph{Foundations of quantum group theory}, Cambridge University
Press, Cambridge, 1995.

\bibitem{PinterPhD}
G.~Pinter, \emph{Epstein-{G}laser renormalization: {F}inite renormalizations,
the {S}-matrix of $\phi^4$ theory and the action principle}, Ph.D. thesis,
Hamburg University, 2000.

\bibitem{PinterHopf}
\bysame, \emph{The {H}opf algebra structure of {C}onnes and {K}reimer in
{E}pstein-{G}laser renormalization}, Lett. Math. Phys. \textbf{54} (2000),
227--233.

\bibitem{Pinter}
\bysame, \emph{Finite renormalization in the {E}pstein-{G}laser framework and
  renormalization of the {S}-matrix of {$\Phi^4$}-theory}, Ann. Phys. 
(Leipzig) \textbf{10} (2001), 333--363.

\bibitem{ReedSimonII}
M.~Reed and B.~Simon, \emph{Methods of modern mathematical physics {II}:
{F}ourier analysis, self-adjointness}, Academic Press, New York, 1975.

\bibitem{Ritter}
W.G. Ritter, \emph{Description of noncommutative theories and
  matrix models by {W}ightman functions}, J. Math. Phys.
\textbf{45} (2004), 4980--5002.

\bibitem{Schmitt87}
W.~Schmitt, \emph{Antipodes and incidence coalgebras}, J. Combin. Theor. A
\textbf{46} (1987), 264--290.

\bibitem{Schmitt94}
\bysame, \emph{Incidence {H}opf algebras}, J. Pure Appl. Alg. \textbf{96}
(1994), 299--330.

\bibitem{vanderLaan}
P.~van~der Laan, \emph{Operads and the {H}opf algebras of renormalisation},
(2003), math-ph/0311013.

\bibitem{MoerdijkvanderLaan}
P.~van~der Laan and I.~Moerdijk, \emph{The renormalisation bialgebra and
operads},  (2002), hep-th/0210226.

\end{thebibliography}
\end{document}